\documentclass[aps,prd,10pt,twocolumn,notitlepage,nofootinbib,superscriptaddress]{revtex4-2}
\usepackage{amsmath,amssymb,amsfonts}
\usepackage{bbm,bm}
\usepackage{graphicx}
\usepackage[utf8]{inputenc}
\usepackage{amsmath,amsfonts,amssymb,graphicx,subfigure}
\usepackage{slashed,color}
\usepackage{multirow}
\usepackage{lineno}
\usepackage{ulem}
\usepackage{verbatim}
\usepackage[svgnames]{xcolor}
\usepackage{hyperref}
\usepackage{amsmath,amssymb,bm}
\usepackage{booktabs}
\usepackage{xcolor}
\usepackage{tikz}
\usepackage[compat=1.1.0]{tikz-feynman}
\usepackage{listings}

\lstdefinestyle{frcode}{
  basicstyle=\ttfamily\scriptsize,
  columns=fullflexible,
  breaklines=true,
  keepspaces=true,
  showstringspaces=false,
  frame=single,
  aboveskip=3pt,
  belowskip=3pt,
  xleftmargin=2pt,
  xrightmargin=2pt
}

\definecolor{qcol}{HTML}{1B1B1B}
\definecolor{zcol}{HTML}{12355B}
\definecolor{wcol}{HTML}{E4572E}
\definecolor{scol}{HTML}{0B7A75}
\definecolor{vcol}{HTML}{12355B}
\definecolor{blobc}{HTML}{9AA5AD}
\definecolor{labc}{HTML}{1B1B1B}

\usepackage{hyperref}
\hypersetup{colorlinks=true, linkcolor=blue, citecolor=ForestGreen,
filecolor=magenta, urlcolor=ForestGreen,
  pdftitle={Boukidi, et al [ArXiv:]},
  pdfauthor={Boukidi, et al},
  pdfsubject={Subject},pdfcreator={Creator},pdfproducer={Producer},
  pdfkeywords={Scotogenic Model}{Hadron Colliders}{Precision QCD Computations}{Extended Scalar Models}
}
\graphicspath{{Figs/}}

\newcommand{\libName}{\texttt{SM\_Scoto}}
\newcommand{\mgamc}{\texttt{mg5amc}}
\newcommand{\py}{\texttt{PY8}}

\def\ab{{\rm\ ab}}
\def\fb{{\rm\ fb}}
\def\pb{{\rm\ pb}}
\newcommand{\invfb}{{\rm ~fb^{-1}}}
\newcommand{\invab}{{\rm ~ab^{-1}}}
\def\eV{{\rm\ eV}}
\def\keV{{\rm\ keV}}

\def\GeV{{\rm\ GeV}}
\def\TeV{{\rm\ TeV}}

\definecolor{magenta}{HTML}{FF00FF}
\definecolor{cornflowerblue}{HTML}{6495ED}
\definecolor{turquoise}{HTML}{40E0D0}

\definecolor{darkgreen}{rgb}{0.0, 0.2, 0.13}
\definecolor{darkmagenta}{rgb}{0.55, 0.0, 0.55}
\definecolor{amber}{rgb}{1.0, 0.6, 0.0}

\newcommand{\confirm}[1]{{\color{black}#1}}

\newcommand{\orcid}[1]{\,\href{https://orcid.org/#1}{\includegraphics[width=9pt]{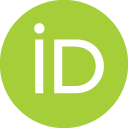}}}
\newcommand{\orcidMB}{0000-0001-9961-8772} %%% MB
\newcommand{\orcidCR}{0000-0003-1267-2202} %%% CR
\newcommand{\orcidRR}{0000-0002-3316-2175} %%% RR

\begin{document}
\leftline{}
\rightline{IFJPAN-IV-2026-17, COMETA-2026-34, MLQC4FC-2026-0005} 

\title{Precise QCD Predictions for the Scotogenic Model at Colliders}

\author{Mohammed Boukidi\ \orcid{\orcidMB}}
\email{mboukidi@ifj.edu.pl}

\author{Camila Ramos\ \orcid{\orcidCR}}
\email{camila.ramos@ifj.edu.pl}

\author{Richard Ruiz\ \orcid{\orcidRR}}
\email{rruiz@ifj.edu.pl}

\affiliation{Institute of Nuclear Physics -- Polish Academy of Sciences {\rm (IFJ PAN)}, ul. Radzikowskiego, Krak{\'o}w, 31-342, Poland}

%%-----------------------------------------------------------------

\begin{abstract}
Motivated by the anticipated data rates 
of the high-luminosity phase of the Large Hadron Collider,
we report the first predictions up to next-to-leading order 
in quantum chromodynamics with parton-shower matching 
for the production and decay 
of scalars and heavy neutrinos 
from the Scotogenic model for neutrino masses and dark matter
at $\sqrt{s}=14$ TeV and $\sqrt{s}=100$ TeV.
We carry out our simulations using the 
\texttt{FeynRules}+\texttt{MadGraph5\_aMC@NLO}+\texttt{Pythia8} pipeline.
As a byproduct, we report the public release of the 
\texttt{SM\_Scoto} Universal \texttt{FeynRules} Object libraries
for modeling the Scotogenic model in high-energy processes.
We comment on prospects of observing new rare $Z$ decays
at future $e^+e^-$ collider facilities as well as 
on the production of new particles at a multi-TeV $\mu^+\mu^-$ collider.
\end{abstract}

\date{\today} 

\maketitle

%%-----------------------------------------------------------------
%%-----------------------------------------------------------------
%%-----------------------------------------------------------------
%%-----------------------------------------------------------------
%%-----------------------------------------------------------------
%%-----------------------------------------------------------------
%%-----------------------------------------------------------------
\section{Introduction}
\label{sec:introduction}

The discovery of nonzero neutrino masses and 
the success of the $3\times3$ paradigm for neutrino 
mixing~\cite{Ahmad:2002jz, Ashie:2005ik} 
constitute one of the most profound challenges to the Standard Model 
of particle physics (SM).
In the SM, right-handed neutrinos $(\nu_R)$ are absent 
and left-handed neutrinos remain massless Weyl fermions.
The model 
must therefore be augmented by new particles and interactions
in order to accommodate neutrino oscillation data~\cite{Ma:1998dn}.

Tree-level models for generating neutrino masses,
e.g., the Types I-III Seesaw models,
remain among the best proposals 
for explaining neutrino masses that are naturally small 
{$m_\nu\lesssim\mathcal{O}(0.5)\eV$}~\cite{KATRIN:2024cdt}
when compared to the electroweak (EW) scale 
$v=\sqrt{2}\langle H\rangle\approx246\GeV$.
For reviews, 
summaries of constraints,
and projections for future sensitivity, 
see Ref.~\cite{Deppisch:2015qwa,Cai:2017mow,FileviezPerez:2022ypk}
and references therein.

Despite their theoretical successes,
tree-level models have disadvantages that 
motivate broader explorations 
for the origin of neutrino masses.
The simplest models, for example, solve the problem
of tiny masses
by hypothesizing the existence
of a new mass scale that is
typically far beyond the EW scale.
In other words,
replace one hierarchy problem with another.
Low-scale variants of these models 
are similarly unsatisfying 
due to a proliferation 
of Yukawa couplings and mass parameters~\cite{Moffat:2017feq}.

Radiative neutrino mass models
address such \textit{theoretical} concerns 
by hypothesizing additional
symmetries,
such as discrete symmetries and parities.
These symmetries forbid neutrino masses at tree level
but allow their generation radiatively.
Tiny neutrino masses are then understood
to originate from
loop quantum corrections
that are suppressed further by
a new mass scale close to the EW scale.
Importantly, 
concerns of radiative stability and fine-tuning, e.g.,
two-loop corrections being
as large as the one-loop result,
are resolvable
by further augmenting the simplest models
by vector-like fermions
or another minimal particle 
content~\cite{Bouchand:2012dx,Merle:2015gea,Merle:2015ica,Merle:2016scw,Lindner:2016kqk,Cai:2017jrq}.

Among the forest of radiative mass 
models~\cite{Cai:2017jrq,Cordero-Carrion:2019qtu} 
is the Scotogenic model~\cite{Ma:2006km,Escribano:2020iqq}.
The model extends the SM 
by a second Higgs doublet $(\Phi_2)$
and
three generations
of right-handed Majorana neutrinos $(\nu_R)$,
but
posits that the new fields
additionally carry an exact $\mathbb{Z}_2$ parity.
The new scalars couple to SM particles 
through EW gauge couplings,
new Higgs-portal couplings,
and new Yukawa couplings,
with the latter two generating 
Majorana neutrino masses at one loop.
And due to the parity, 
the lightest particle
in the spectrum is stable 
and a candidate for dark matter~\cite{Ma:2006km,AristizabalSierra:2008cnr,Molinaro:2014lfa,Merle:2016scw,Bernal:2017kxu,Okada:2026upm,Wang:2026ytg}.

Motivated by the upcoming high-luminosity (HL) phase 
of the Large Hadron Collider (LHC),
in this work we revisit the predictions
for the production and decay of particles 
in the Scotogenic model at colliders.
We complement previous investigations 
of the Scotogenic model at the 
LHC~\cite{AristizabalSierra:2008cnr,Hessler:2014ssa,Hessler:2016kwm,Ahriche:2017iar,Baumholzer:2019twf,Ardila-Tafurth:2025qdf,Wang:2026miy,Wang:2026wid}
by reporting, for the first time, 
the total and differential cross sections 
for a variety of production channels
in hadronic collisions at $\sqrt{s}=14\TeV$ and $\sqrt{s}=100\TeV$,
up to next-to-leading order (NLO) in quantum chromodynamics (QCD)
with matching to a parton shower (PS).

As the model of Ref.~\cite{Ma:2006km}
serves as a baseline
for a large class of radiative models
and due to the present lack of certainty 
on the absolute values and nature of neutrino masses,
we take a phenomenological approach 
and treat the production and decay of Scotogenic particles 
independent of a particular underlying model for lepton flavor.
As a byproduct, we also report the public release  
of the {\libName} libraries,
a set of Universal \texttt{FeynRules} Output (UFO) 
libraries~\cite{Degrande:2011ua,Alloul:2013bka,Darme:2023jdn}
that allow one to simulate high-energy processes
in the Scotogenic model 
using contemporary, high-energy software environments.

The remainder of this work continues in the following order:
In Sec.~\ref{sec:theory}, we summarize the Scotogenic model 
and our phenomenological parameterization of the framework.
Current experimental constraints are also summarized there.
In Sec.~\ref{sec:setup}, we describe our computational setup
for our numerical studies, 
including the creation of the {\libName} libraries.
In Sec.~\ref{sec:decay}, we survey decay rates and lifetimes of 
Scotogenic particles.
This also serves as validation check of our computational setup.
In Sec.~\ref{sec:production}, 
we report production-level cross sections 
at the total and differential level 
for the $\sqrt{s}=14\TeV$ LHC 
and a hypothetical $\sqrt{s}=100\TeV$ $pp$ collider.
In Sec.~\ref{sec:outlook}, 
we give an outlook for explorations 
at future $e^+e^-$ and $\mu^+\mu^-$ colliders.
We conclude in Sec.~\ref{sec:conclusion}.
Additional details of our computational setup 
are reported in Apps.~\ref{sec:ufo} and \ref{sec:mg5}.

%%-----------------------------------------------------------------
%%-----------------------------------------------------------------
\section{The Phenomenological Scotogenic Model}
\label{sec:theory}

The Scotogenic model~\cite{Ma:2006km} 
and its extensions have been studied 
extensively. We refer readers to 
Refs.~\cite{Toma:2013zsa,Vicente:2014wga,Cai:2017jrq,Cai:2017mow,Escribano:2020iqq}
and references therein 
for detailed discussions of the literature.
In this section, we summarize only the most essential 
ingredients of our phenomenological parameterization 
of the model relevant for our study.

The Scotogenic model extends the SM 
by a scalar SU$(2)_L$ doublet $\eta$
and three, right-handed\footnote{Chiral fermions 
are denoted by $\psi_{L/R}=P_{L/R}\psi$ with the 
usual chiral projection operators defined 
as $P_{L/R}=(1\mp\gamma^5)/2$.} 
Majorana singlets fields  
$(\nu_R)_a$, with $a=1,2,3$.
The new fields are odd under an exact $\mathbb{Z}_2$ symmetry, 
while SM fields are even under the parity. 
Table~\ref{tab:fields} gives the gauge quantum
numbers and $\mathbb{Z}_2$ charges of the model's 
scalar and lepton sectors. 

%%-----------------------------------------------------------------
\subsection{Lagrangian and Particle Spectrum}
\label{sec:theory_lag}

The full Lagrangian of the model is given by~\cite{Ma:2006km} 
\begin{align}
\label{eq:lagFull}
\mathcal{L}_{\rm Scoto} &= \mathcal{L}_{\rm SM}
+ \mathcal{L}_{\rm Kin} 
+ \mathcal{L}_{\rm Yuk}
- \mathcal{V}(H,\eta) 
+ \mathcal{L}_{\nu}^{\rm 1-loop}\ .
\end{align}
$\mathcal{L}_{\rm SM}$ is SM Lagrangian, 
excluding its scalar potential $\mathcal{V}_{\rm H}$.
$\mathcal{L}_{\rm Kin}$ is the kinetic term 
for $\eta$ and $(\nu_R)_a$, and is given by 
\begin{align}
\label{eq:lag_kin}
\mathcal{L}_{\rm Kin} &= (D_\mu\eta)^\dagger(D^\mu\eta)
+ \frac{i}{2}\overline{(\nu_R^c)_a} \!\not\!\partial (\nu_R)_b  
- \frac{\mu_R^{ab}}{2}\overline{(\nu_R^c)_a}(\nu_R)_b
\nonumber\\
=&\ (D_\mu\eta)^\dagger(D^\mu\eta)
+ \frac{i}{2}\overline{(N_k^c)_k} \!\not\!\partial N_k
- \frac{M_{N_k}}{2}\overline{N^c_k}N_k\ .
\end{align}
$D_\mu = [\partial_\mu 
+ i g_W \hat{T}_L^i W_\mu^i + i (g_Y/2) \hat{Y} B_\mu]$ 
is the EW covariant derivative of $\eta$,
$\mu_R^{ab}$ are right-handed Majorana masses 
in the chiral basis,
and, after a trivial rotation, 
$M_{N_k}$ are the mass eigenvalues 
for the mass eigenstates $N_1,N_2,N_3$.

The full scalar potential of the model is 
\begin{subequations}
\begin{align}
\label{eq:potential}
\mathcal{V}(H,\eta)\ &=\ \mathcal{V}_H +\mathcal{V}_\eta(H,\eta),
\\
\mathcal{V}_H\ &=\ \mu_H^2H^\dagger H
+\frac{\lambda_1}{2}(H^\dagger H)^2\ ,
\end{align}
where the scalar doublets are written in the $R_\xi$ gauge as
\begin{equation}
	H=
	\begin{pmatrix}
		-iG^+\\[1mm]
		\frac{v+h+iG^0}{\sqrt{2}}
	\end{pmatrix},
	\qquad
	\eta=
	\begin{pmatrix}
		-i\eta^+\\[1mm]
		\frac{\eta_R+i\eta_I}{\sqrt{2}}
	\end{pmatrix}.
	\label{eq:doublets}
\end{equation}
Assuming charge-parity (CP) conservation\footnote{A rephasing of 
$\eta$ allows us to take the dimensionless parameter $\lambda_5$
to be real without a loss of generality~\cite{Ma:2006km}.},
the remaining terms of the full scalar potential 
are given by
\begin{align}
	\mathcal{V}_{\rm \eta}(H,\eta)\ 
	&=\ 
	\mu_\eta^2\eta^\dagger\eta
	+\frac{\lambda_2}{2}(\eta^\dagger\eta)^2
	+\lambda_3(H^\dagger H)(\eta^\dagger\eta)
	\nonumber\\
	+\ &
	\lambda_4(H^\dagger\eta)(\eta^\dagger H)
	+\frac{\lambda_5}{2}
	\left[(H^\dagger\eta)^2+\mathrm{H.c.}\right]\ .
	\label{eq:eta-potential}
\end{align}
\end{subequations}

For $\mu_H^2<0$ and $\mu_\eta^2>0$, 
only $H$ acquires a vacuum expectation value 
at\footnote{Note that Ref.~\cite{Ma:2006km} adopts the convention 
$v= \langle H\rangle$.} 
$v=\sqrt{2}\langle H\rangle\approx 246\GeV$.
Since $\langle\eta\rangle=0$,
the $\mathbb{Z}_2$ symmetry remains exact.
While renormalization group running can 
spontaneously break $\mathbb{Z}_2$ symmetry,
introducing additional particles can stabilize the 
running~\cite{Merle:2015gea,Merle:2015ica,Merle:2016scw,Lindner:2016kqk}.
Assuming the $\mathbb{Z}_2$ symmetry holds, 
none of the states in $H$ and $\eta$ mix,
and the lightest $\mathbb{Z}_2$-odd state is stable.
If also electrically neutral, 
the lightest $\mathbb{Z}_2$-odd state
is a candidate for particle dark matter.

\begin{table}[!t]
	\caption{Gauge quantum numbers and $\mathbb{Z}_2$ charges 
    of the fields relevant to the Scotogenic sector
    for $k=1,2,3$.}
	\label{tab:fields}
	\setlength{\tabcolsep}{3.2pt}
	\begin{ruledtabular}
		\begin{tabular}{lccccc}
			Field & Spin & $SU(3)_c$ & $SU(2)_L$ & $Y$ & $\mathbb{Z}_2$ \\
			$H$        & $0$        & $\mathbf{1}$ & $\mathbf{2}$ & $+\tfrac12$ & $+$ \\
			$L_k$ & $\tfrac12$ & $\mathbf{1}$ & $\mathbf{2}$ & $-\tfrac12$ & $+$ \\
			$\eta$     & $0$        & $\mathbf{1}$ & $\mathbf{2}$ & $+\tfrac12$ & $-$ \\
			$(\nu_R)_k/N_k$      & $\tfrac12$ & $\mathbf{1}$ & $\mathbf{1}$ & $0$         & $-$
		\end{tabular}
	\end{ruledtabular}
\end{table}

After EW symmetry breaking and in the mass basis,
we can identify $G^{\pm},G^0$ in Eq.~\eqref{eq:doublets} 
as the EW Goldstone bosons,
$h$ as the SM Higgs boson with $m_h\approx125\GeV$,
$\eta^\pm$ as new charged scalar fields,
$\eta_R$ as an electrically neutral scalar,
and $\eta_I$ as an electrically neutral pseudoscalar.

In terms of the $\{\lambda_i\}$, 
the masses of $\eta^\pm$, $\eta_{R}$, and $\eta_{I}$ are
\begin{subequations}
	\label{eq:scalar-masses}
\begin{align}
	m_{\eta^\pm}^2
	&=
	\mu_\eta^2+\frac{\lambda_3v^2}{2}\ ,
	\\
	m_{\eta_R}^2
	&=
	\mu_\eta^2+\frac{\lambda_Lv^2}{2}\ ,\ 
    \lambda_L=\lambda_3+\lambda_4+\lambda_5,
	\\
	m_{\eta_I}^2
	&=
	\mu_\eta^2+\frac{\lambda_Sv^2}{2}\ ,\ 
    \lambda_S=\lambda_3+\lambda_4-\lambda_5,
\end{align}
with the neutral-scalar mass splitting given by
\begin{align}
	m_{\eta_R}^2-m_{\eta_I}^2=\lambda_5v^2.
	\label{eq:neutral-splitting}
\end{align}
\end{subequations}

The Yukawa interaction that couples 
the new Majorana neutrino fields $\nu_R/N$ 
to SM leptons is given by
\begin{align}
%----------------------------------
	\mathcal{L}_{\rm Yuk}\ =\
    &- \widetilde{Y}_{a b}\,
	\overline{L_a}\,
	\widetilde{\eta}\,
	P_R\ (\nu_R)_b
	+\ 
    \mathrm{H.c.}
%----------------------------------    
\\
    =\
	&-\frac{Y_{\alpha k}}{\sqrt2}\
	\overline{\nu_\alpha}
	(\eta_R-i\eta_I)P_RN_k
	\nonumber\\ 
    &+\
	iY_{\alpha k}\
	\overline{\ell_\alpha}\eta^-P_RN_k\
	+\ 
    \mathrm{H.c.}
	\label{eq:yukawa-expanded}
\end{align}
$\widetilde{\eta}=i\sigma_2\eta^*$ 
is the conjugate of the $\eta$ doublet 
rotated in SU$(2)_L$ space.
$L_a=((\nu_L)_a,(l_L)_a)^T$ is the usual SM lepton doublet  
in the interaction basis, 
with $a=1,2,3$ running over generations.
We work in the basis where the SM
charged-lepton Yukawa matrix is diagonal, 
and in Eq.~\eqref{eq:yukawa-expanded}
(trivially) rotate the leptons into the flavor basis.
After absorbing both leptonic rotation matrices,
the Yukawa couplings $Y_{\alpha k}$ 
run over lepton flavors $\alpha=e,\mu,\tau$
and singlet neutrino mass states $k=1,2,3$.

%%-----------------------------------------------------------------
\subsection{Neutrino Masses}
\label{sec:theory_nu}

\begin{figure}[!t]
	\centering
	\includegraphics[width=.75\columnwidth]{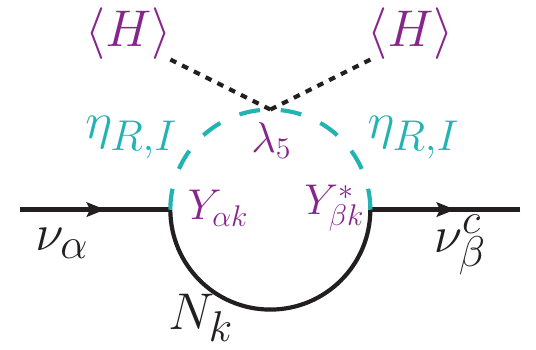}
	\caption{Diagram that generates  
    neutrino masses at one loop in the Scotogenic model.
    Drawn using \texttt{JaxoDraw2}~\cite{Binosi:2008ig}.}
	\label{fig:scotoLHC_nuMass1Loop}
\end{figure}

The $\mathbb{Z}_2$ symmetry forbids
Dirac Yukawa couplings of the form $\overline L\widetilde H\nu_R$.
Active neutrinos are therefore massless at tree level 
but acquire left-handed Majorana masses 
at one loop through the SM neutrinos' Yukawa couplings
to $\eta$ [Eq.~\eqref{eq:yukawa-expanded}]
and the $\eta$'s couplings 
to the neutral component of the SM Higgs field $H$
[Eq.~\eqref{eq:eta-potential}].
This interaction with the relevant couplings is illustrated in Fig.~\ref{fig:scotoLHC_nuMass1Loop}.

In the flavor basis with indices $\alpha,\beta=e,\mu,\tau$,
the neutrino mass matrix at one loop is exactly~\cite{Ma:2006km}
\begin{align}
	&(\mathcal{M}_\nu^{\rm 1-loop})_{\alpha\beta}\
	=\
	\sum_{k=1}^3
	\frac{Y_{\alpha k}Y_{\beta k}^*M_{N_k}}{16\pi^2}
    \nonumber\\
    \times&
	\Big[
    \frac{m_{\eta_R}^2}{m_{\eta_R}^2-M_{N_k}^2}\log\frac{m_{\eta_R}^2}{M_{N_k}^2}
    -
    \frac{m_{\eta_I}^2}{m_{\eta_I}^2-M_{N_k}^2}\log\frac{m_{\eta_I}^2}{M_{N_k}^2}
    \Big]
	\label{eq:neutrino-mass}
\end{align}
For $\lambda_5=0$, $\eta_R$ and $\eta_I$ are mass-degenerate 
and the two terms cancel. 
A nonzero mass splitting between the neutral scalars 
$\eta_R$ and $\eta_I$ is therefore required to 
generate nonzero neutrino masses.
While the observation of two mass splittings 
among the light neutrinos~\cite{Esteban:2024eli} 
requires at least two generations of $N_k$,
we retain all three $N_k$.

Due to the $\mathbb{Z}_2$ parity 
none of the $N_k$ mix with the light neutrinos. 
As a result, the decomposition 
of neutrinos' flavor states $\nu_\alpha$ 
in terms mass states $\nu_k$ is governed by
\begin{align}
\label{eq:numixing}
(\nu_L)_\alpha\ &=\ 
\sum_{k=1}^3\ U_{\alpha k}^{\rm PMNS}\ \nu_k\     
\end{align}
where $U_{\alpha k}^{\rm PMNS}$ is the familiar 
Pontecorvo-Maki-Nakagawa-Sakata (PMNS) mixing matrix
measured by neutrino oscillation experiments.
In other words, neutrino oscillations are described 
by the standard $3\times3$ PMNS paradigm.
Consequentially, we can rotate 
$(\mathcal{M}_\nu^{\rm 1-loop})_{\alpha\beta}$
into the mass basis using $U^{\rm PMNS}$ 
to obtain the mass matrix for light neutrinos 
in the mass basis,
\begin{align}
\mathcal{L}_{\nu}^{\rm 1-loop} &= 
\frac{1}{2}
\overline{(\nu_L)_\alpha}\ 
(\mathcal{M}_\nu^{\rm 1-loop})_{\alpha\beta}\ 
(\nu_L^c)_\beta 
\\
&= 
\frac{1}{2}
\overline{\nu_k} 
\left[(U_{k\alpha}^{\rm PMNS})^*
(\mathcal{M}_\nu^{\rm 1-loop})_{\alpha\beta} 
U_{\beta k}^{\rm PMNS}\right]
\nu_k^c 
\\
&=\frac{1}{2}
\overline{\nu_k}\
{\rm diag}(m_{\nu_k})\  
\nu_k^c\ ,
\label{eq:lag_numass}
\end{align}
where ${\rm diag}(m_{\nu_k})$ is the mass matrix in the mass basis.

%%-----------------------------------------------------------------
\subsection{A Phenomenological Approach to Discovery}
\label{sec:theory_pheno}

An important consequence of Eq.~\eqref{eq:numixing}
is that the mass matrix in the flavor basis 
[Eq.~\eqref{eq:neutrino-mass}], 
and hence Yukawa couplings of Eq.~\eqref{eq:yukawa-expanded}, 
can be caste in terms 
of known oscillation parameters~\cite{Esteban:2024eli}
by the 
relationship~\cite{Hehn:2012kz,Toma:2013zsa,Vicente:2014wga}
\begin{align}
\label{eq:mass_osc}
U_{\alpha k}^{\rm PMNS} 
\left[{\rm diag}(m_{\nu_k})\right]_{kk} 
(U_{k\beta }^{\rm PMNS})^*  =
    (\mathcal{M}_\nu^{\rm 1-loop})_{\alpha\beta}\ .
\end{align}
As discussed in Sec.~\ref{sec:decay},
these Yukawa couplings are encoded 
in branching rates and branching ratios 
of Scotogenic particles.
This demonstrate a synergy 
among collider experiments,
dedicated facilities to search for lepton flavor violation,
neutrinoless $\beta\beta$ decay experiments,
dark matter search experiments,
and neutrino oscillation experiments 
in studying the neutrino and dark sectors.

The goal of this work is to explore 
the collider phenomenology 
of the simplest (canonical) Scotogenic model 
in order to provide broad guidance 
on testing the Scotogenic model 
at the LHC and other accelerator facilities.
Therefore, 
for purposes of discoverying new phenomena at colliders,  
we neglect Eq.~\eqref{eq:mass_osc}
since the production of Scotogenic particles must 
generally (but not exclusively) go through SM gauge couplings 
(e.g., $u\overline{u}\to \gamma^*/Z^*\to \eta^+\eta^-$)
or through the decays of particles produced 
via SM gauge couplings (e.g., $Z\to N_1N_1\nu\overline{\nu}$).
If particles consistent with the Scotogenic model are discovered 
at colliders, then Eq.~\eqref{eq:mass_osc} 
will be essential to falsifying competing models 
for neutrino masses 
as inclusive production rates at colliders are insufficient 
to overcome the variability of scalar and Yukawa couplings~\cite{Hehn:2012kz,Toma:2013zsa,Vicente:2014wga,delAguila:2013yaa,Ruiz:2022sct}.

As a consequence of neglecting Eq.~\eqref{eq:mass_osc}, 
we treat 
the scalar masses $m_\eta$ in Eq.~\eqref{eq:scalar-masses}
and the Yukawa couplings $Y_{\alpha k}$
in Eq.~\eqref{eq:yukawa-expanded} as free parameters
that can be constrained independently of one another.
We also treat light neutrinos as massless particles,
with flavor states $\nu_\alpha$ being distinct 
from the antiparticle states $\overline{\nu_\alpha}$.
All three $N_k$ are treated as Majorana fermions.
This gives us the ability to make predictions 
for the Scotogenic model that are agnostic 
of a particular underlying flavor symmetry.

%%-----------------------------------------------------------------
\subsection{Constraints}
\label{sec:theory_constraints}

We now summarize theoretical and experimental constraints 
on the Scotogenic model.

\begin{itemize}
    \item \textbf{Vacuum Stability and Perturbativity:} 
    The new terms in the scalar potential $\mathcal{V}(H,\eta)$ 
    are required to maintain stability of the EW vacuum, which is achieved by imposing, at tree-level~\cite{Ahriche:2017iar}:
\begin{equation}
    \lambda_{1,2,3} >0, \quad \lambda_{3} + \lambda_{4} - |\lambda_{5}| > -2 \sqrt{\lambda_{1}\lambda_{2}}.
\end{equation}
Additionally, these quartic couplings and the Yukawa interactions must remain within the perturbative regime, $|\lambda_{i}|, |Y_{\alpha k}| < 4 \pi$.

\item \textbf{Parity Problem:} 
For sufficiently large right-handed neutrino masses,
renormalization group running can spontaneously break the 
exact $\mathbb{Z}_2$ parity~\cite{Merle:2015gea,Merle:2015ica}. 
This requires at the scale $\mu_0=M_Z$
\begin{align}
 m_\eta^2\ \gtrsim\ \max(M_{N_1},M_{N_2},M_{N_3})
\end{align}
or hypothesizing additional SU$(2)_L$ multiplets~\cite{Merle:2016scw}.

\item \textbf{Oblique Parameters:} The $\eta$ doublet 
can potentially contribute to the S and T oblique parameters~\cite{Arcadi:2019lka, Ahriche:2017iar}, which are strongly constrained by LEPI~\cite{ALEPH:2005ab}. The most recent global fit of the oblique parameters to EW precision data~\cite{ParticleDataGroup:2026mpi} results in
\begin{subequations}
\begin{align}
    S|_{U=0} = 0.008 \pm 0.071, 
    \\ 
    \quad T|_{U=0} = 0.021 \pm 0.055.
\end{align}
\end{subequations}

\item \textbf{LEP Collider Constraints:} 
Precise measurements on the $Z$ and $W$ decay widths at LEP 
rule out the bosons decaying directly to $\eta_1\eta_2$ pairs,
and require that~\cite{Cao:2007rm,Lundstrom:2008ai,Avila:2025qsc,Belyaev:2016lok} 
\begin{subequations}
\begin{align}
 m_{\eta_{R}} + m_{\eta_{I}}\ &>\ M_{Z}\ ,
 \\
 2 m_{\eta^{\pm}}\ &>\ M_{Z}\ ,
 \\
 m_{\eta_{R,I}} + m_{\eta^{\pm}}\ &>\ M_{W}\ .
\end{align}
Searches for charginos and neutralinos have been reinterpret~\cite{Pierce:2007ut} to convert to the exclusion of
\begin{align}
 m_{\eta^{\pm}}\ <\ 70-90\GeV\ \text{at 95\% C.L.}\ ,
\end{align}
\end{subequations}
while searches for neutral scalars can also further constrain the parameter space for $m_{\eta_{R}}$, depending on the mass splitting~\cite{Pierce:2007ut}.

\item \textbf{LHC Constraints:} Searches for neutral scalar pairs 
decaying to electrons and muons in $pp$ collisions 
at $\sqrt{s}=13-13.6\TeV$ with $\mathcal{L}=35-138\invfb$
by the CMS experiment exclude at 95\% C.L.~\cite{CMS:2026cjr}
\begin{subequations}
\begin{align}
m_{\eta_R} < 108\GeV\ &\text{for}\ m_{\eta_R}-m_{\eta_I} = 78\GeV\ ,
\\
m_{\eta_R} < 70\GeV\ &\text{for}\ m_{\eta_R}-m_{\eta_I} = 40-90\GeV .
\end{align}
\end{subequations}

\item \textbf{Effective Majorana Mass:}
A search for nuclear neutrinoless $\beta\beta$ decay 
in germanium by the LEGEND Collaboration 
places a limit of~\cite{LEGEND:2025jwu}
\begin{subequations}
\begin{align}
    m_{ee}\ &<\ 75-200\ {\rm meV}\ \text{at\ 90\% C.L.}
\end{align}
on the $(\alpha,\beta)=(e,e)$ element 
of $(\mathcal{M}_\nu^{\rm 1-loop})_{\alpha\beta}$.
Searches for lepton number violation 
in the  $W^\pm W^\pm\to\ell_1^\pm\ell_2^\pm$
scattering channel at the LHC place the analogous 
bounds at 95\% CL~\cite{CMS:2022hvh,ATLAS:2023tkz,ATLAS:2024rzi}
\begin{align}
m_{ee},\  
m_{e\mu},\  
m_{\mu\mu}\ &\ < 10.8-24\GeV\ .
\end{align}
\end{subequations}

\item \textbf{Searches for Dark Matter:}
Searches for dark matter in liquid xenon
by the LZ experiment have yielded one event 
after an exposure of 2.84 ton-years  
with a nuclear recoil of  
$\Delta E_{\rm recoil}=248\pm23({\rm stat})\pm23({\rm sys})\keV$
with $2.6\sigma$ confidence~\cite{LZ:2026axp}.
This is consistent with 
$m_{\eta_R}\in [691,1479]\GeV$ 
and $m_{\eta_I}-m_{\eta_R}=[360,369]\keV$\cite{Okada:2026upm,Wang:2026ytg}.
We refer to Refs.~\cite{Avila:2025qsc, Arcadi:2024ukq, Longas:2026yxq} for overviews 
on limits derived from other direct detection experiments 
for the Scotogenic model and some of its variations. 

\end{itemize}

%-----------------------------------------------------------------
%-----------------------------------------------------------------
%-----------------------------------------------------------------
%-----------------------------------------------------------------
%-----------------------------------------------------------------
\section{Computational Setup and the {\libName} UFO Libraries}
\label{sec:setup}

In order to carry out
our numerical calculations,
we implemented the Scotogenic model
Lagrangian as given in Eq.~\eqref{eq:lagFull}
into \texttt{FeynRules}
(v2.3.49)~\cite{Christensen:2008py,Alloul:2013bka},
adapting the default
\texttt{sm.fr} file (v1.4.7)
for the SM part of the Lagrangian.
As discussed in Sec.~\ref{sec:theory_pheno}
we omit $\mathcal{L}_\nu^{\rm 1-loop}$.

QCD ultraviolet and $R_2$ counter terms
up to $\mathcal{O}(\alpha_s)$
were extracted using
\texttt{NLOCT}
(v1.02)~\cite{Degrande:2014vpa}
and
\texttt{FeynArts}
(v3.11)~\cite{Hahn:2000kx}.
Feynman rules that are
accurate at tree level in EW couplings
and up to one loop in the strong coupling
were then packaged into a series of
Universal \texttt{FeynRules} Output
(UFO)~\cite{Degrande:2011ua}
libraries that we collectively
call the {\libName} UFO
libraries\footnote{Individual UFO libraries,
their variants, and the associated \texttt{FeynRules}
model files are all publicly available
from the URL \href{https://github.com/FeynRules/Models/tree/main/SM_Scoto}{https://github.com/FeynRules/Models/tree/main/SM\_Scoto}.}.

Matrix elements for various processes
were generated and evaluated
using the simulation framework
\texttt{MadGraph5\_aMC@NLO} (\texttt{mg5amc})
(\confirm{v3.7.2})~\cite{Stelzer:1994ta,Alwall:2014hca}.
This framework numerically simulates
fully differential processes by employing
helicity amplitudes
in the \texttt{HELAS}
basis~\cite{Hagiwara:1985yu,Murayama:1992gi},
\texttt{MadLoop}~\cite{Hirschi:2011pa,Hirschi:2015iia}
for virtual radiative contributions,
and the MC@NLO
formalism~\cite{Frixione:2002ik}
as implemented in \texttt{MadFKS}~\cite{Frixione:1995ms,Frixione:1997np,Frederix:2009yq}
for real radiative contributions.
The Feynman rules encoded
in the {\libName} UFO are
adapted to \texttt{HELAS}
using the built-in \texttt{ALOHA} module~\cite{deAquino:2011ub}.

Events are parton showered using
\texttt{Pythia}
(\py)
(v8.306)~\cite{Bierlich:2022pfr},
with underlying event,
``primordial'' quantities,
and electromagnetic (QED) showering
all enabled.
The HEPMC output~\cite{Dobbs:2001ck}
of {\py} is analyzed directly
using a custom analysis
libraries\footnote{Analysis scripts
are publicly available from the URL\\
\href{https://gitlab.cern.ch/riruiz/public-projects/-/tree/master/ScotoLHC}{gitlab.cern.ch/riruiz/public-projects/-/tree/master/ScotoLHC}\ .}
based on the open-sourced \texttt{pyhepmc}
project~\cite{Buckley:2019xhk,hans_dembinski_2022_7013498}.

For additional technical details
on the implementation of the Scotogenic
model into the {\libName} UFO libraries,
see App.~\ref{sec:ufo}.
App.~\ref{sec:mg5}
provides additional information
on our usage of
{\libName}+{\mgamc}+{\py}  pipeline.

%%-----------------------------------------------------------------
%%-----------------------------------------------------------------
\subsection{SM Inputs}
\label{sec:setup_sm}

We fix our SM inputs to those values
listed in the 2026 edition of the
Particle Data Group~\cite{ParticleDataGroup:2026mpi},
\begin{align}
 \alpha_{\rm EW}^{-1}(M_Z) &= 127.955\ ,\
 M_Z \ =91.1879\GeV\ ,
\nonumber\\
G_{\rm F} &=\
1.1663785\times10^{-5}\GeV^{-2}\ ,
\nonumber\\
 \Gamma_Z &= 2.4955\GeV\ ,\
 \Gamma_W = 2.14\GeV\ ,
 \nonumber\\
 m_t(m_t) &=\ 172.60\GeV\ ,\
 m_\tau =\ 1.77693\GeV\ .
 \label{eq:sm_inputs}
\end{align}
By default, we work with $n_f=5$ quarks flavors
and a diagonal Cabibbo-Kobayashi-Maskawa (CKM) matrix
equal to the identity matrix $V^{\rm CKM} = \mathbb{I}_3$.
This corresponds to using the \texttt{SM\_Scotogenic\_NLO} UFO.
For computations where the $\tau$ lepton mass is relevant 
or when using the $n_f=4$ flavor scheme, 
we set $m_b(m_b) =\ 4.186\GeV$ and employ the 
\texttt{SM\_Scotogenic\_MassiveLeptons\_4fs\_NLO} UFO.

For hadronic cross sections,
we use the NNPDF 4.0 QCD NLO + QED NLO
parton density functions (PDFs)
(\texttt{lhaid=335900}) set~\cite{NNPDF:2024djq}
with $\alpha_s(M_Z)=0.1180$.
The PDF employs the LUXqed formalism
for the (inelastic) photon PDF of the 
proton~\cite{Manohar:2016nzj,Manohar:2017eqh}.

We set the factorization $(\mu_f)$ and renormalization $(\mu_r)$ scales to be half the sum of transverse energies of final-state particles
(\texttt{dynamical\_scale\_choice=3}):
\begin{subequations}
\begin{align}
\label{eq:scale_central}
 \mu_f, \mu_r &= \zeta \times \mu_0, \quad\text{where}\quad \zeta=1\quad \text{and}\\
 \mu_0 &= \frac{1}{2} \sum_{f\in\{\text{final state}\}} \sqrt{m_f^2 + p_{Tf}^2}\ .
\end{align}
\end{subequations}
A 9-point scale uncertainty is obtained by varying $\zeta$ over the discrete range $\zeta\in\{0.5,1.0,2.0\}$.
PDF uncertainties are obtained through reweighting 
simulated events over an ensemble of PDF 
replicas~\cite{NNPDF:2024djq}.
Scale and PDF evolution are handled
using \texttt{LHAPDF} (v6.5.5).

%%-----------------------------------------------------------------
%%-----------------------------------------------------------------
\subsection{Scotogenic Inputs}
\label{sec:setup_scoto}

For Scotogenic inputs,
we take the masses of the $\eta$ and $N_k$
fields to be external parameters.
Unless stipulated otherwise,
we fix masses and couplings to be
\begin{align}
\label{eq:scoto_inputs}
 m_{N_2}, m_{N_3}\ &=\ 10^{10}\GeV\ ,\
 Y_{\ell N_1} =\ \delta_{\ell\tau}\ ,
 \nonumber\\
 m_{\eta} &\equiv m_{\eta^\pm} = m_{\eta_R} = m_{\eta_I}\ .
\end{align}
In other words,
for the purposes of this work
and unless specified,
we decouple the particles $N_2$ and $N_3$
(in the sense of
Ref.~\cite{Appelquist:1974tg}),
set all $\eta$ masses equal
(denoted by $m_{\eta}$),
and focus on the couplings
to the $\tau$-flavor sector.

%%-----------------------------------------------------------------
%%-----------------------------------------------------------------
%%-----------------------------------------------------------------
%%-----------------------------------------------------------------
%%-----------------------------------------------------------------
\section{Decay widths}
\label{sec:decay}

From the Lagrangians in Sec.~\ref{sec:theory_lag} 
we can extract the Feynman rules of the model directly 
and compute the decay rates 
for the $\eta$ and $N$ fields.
This also serves as a check 
of our numerical implementation 
of the model.

%%-----------------------------------------------------------------
%%-----------------------------------------------------------------
\subsection{Scalar Decays to Majorana Neutrinos}

Assuming the mass hierarchy 
$m_{\eta^\pm},\ m_{\eta_{R,I}}>M_{N_k}+m_{\ell_\alpha}$,
the decay rates for the processes
\begin{align}
    \eta^\pm \to \ell_\alpha^\pm\ N_k\
    \quad\text{and}\quad
    \eta_{R,I} \to \nu_\alpha\ N_k\ ,
\end{align}    
are given at lowest order by
\begin{align}
%-----------------------------------------------
\Gamma(\eta^+\to\ell_\alpha^+ N_k)
&=\
\frac{m_{\eta^\pm}}{16\pi}\ |Y_{\alpha k}|^2\ 
\lambda_K^{1/2}
\left(1,
\frac{m_{\ell_\alpha}^2}{m_{\eta^\pm}^2},
\frac{M_{N_k}^2}{m_{\eta^\pm}^2}
\right)
\nonumber\\
&\qquad\ \times\
\left(
1-\frac{m_{\ell_\alpha}^2}{m_{\eta^\pm}^2}
-\frac{M_{N_k}^2}{m_{\eta^\pm}^2}
\right)\ ,
\label{eq:charged-width}
%-----------------------------------------------
%-----------------------------------------------
\\
\Gamma(\eta_R\to\nu_\alpha N_k)
&=\
\frac{m_{\eta_R}}{32\pi}\ 
    |Y_{\alpha k}|^2\
	\left(1-\frac{M_{N_k}^2}{m_{\eta_R}^2}
	\right)^2\ .
\label{eq:neutral-width}
%-----------------------------------------------
\end{align}
Here, we use the kinematic K\"all\'en function
\begin{align}
	\lambda_K(x,y,z)\
	&=\ (x-y-z)^2 - 4yz\
    \nonumber\\
    &=\
	x^2+y^2+z^2-2xy-2xz-2yz\ .
	\label{eq:kallen}
\end{align}
These decay widths were previously reported 
in Refs.~\cite{Hessler:2016kwm,Wang:2026wid}
but under the assumption that $m_{\ell_\alpha}\to0$. 

Under CP symmetry, assuming the appropriate mass hierarchies, 
and with the appropriate exchanges of mass ratios in $\lambda_{K}$, 
we also have the decay widths 
\begin{align}
    \Gamma(\eta^-\to\ell_\alpha^- N_k)\ &=\ 
    \Gamma(\eta^+\to\ell_\alpha^+ N_k)\ ,
    \label{eq:charged-width-other}    
    \\
    \Gamma(\eta_R\to\overline{\nu_\alpha} N_k)\ &=\ 
    \Gamma(\eta_I\to\nu_\alpha N_k)\ =\
    \Gamma(\eta_I\to\overline{\nu_\alpha} N_k)
    \nonumber\\
    &=\ \Gamma(\eta_R\to\nu_\alpha N_k)\ .
\label{eq:neutral-width-other}    
\end{align}

\begin{figure}[!t]
	\centering
	\includegraphics[width=\columnwidth]{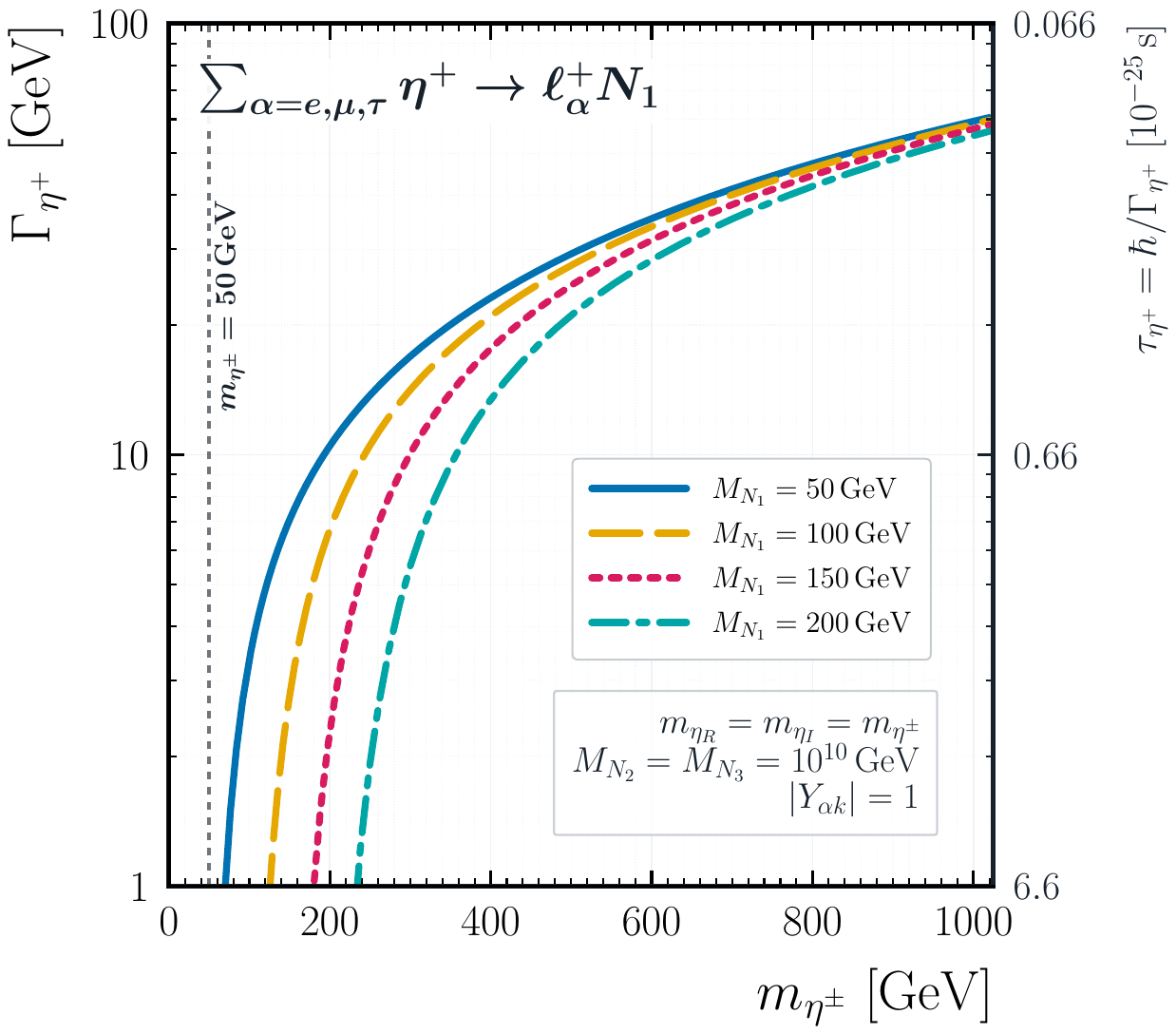}
	\caption{The total width $\Gamma_{\eta^+}$ [GeV]
    as defined in Eq.~\eqref{eq:width_total_eta}
    as a function of scalar mass $m_{\eta^\pm}$ [GeV] 
    for the representative Majorana neutrino masses 
    $M_{N_1}=50$ ({\color{blue}solid}), 
    $100$ ({\color{Goldenrod}dash}), 
    $150$ ({\color{magenta}dot}), 
    and $200\GeV$ ({\color{teal}dash-dot}),
    with $|Y_{\alpha k}|=1$.
	}
	\label{fig:etaC_width}
\end{figure}

\begin{table}[t!]
\centering
%\resizebox{\columnwidth}{!}{
\begin{tabular}{c c c c}
\hline\hline
$m_\eta$
& $\Gamma_{\eta^+\to\tau^+ N_1}$
& $\Gamma_{\eta_R\to\nu_\tau N_1}$
& $\Gamma_{\eta_I\to\nu_\tau N_1}$
\\
{[GeV]} & [GeV] & [GeV] & [GeV]
\\
\hline
%--------------------------------- 
    10  &  $99.5\times10^{-3}$     & $56.0\times10^{-3}$ & $56.0\times10^{-3}$      \\
%--------------------------------- 
	100  & 1.98  & 0.990  & 0.990  \\
%---------------------------------     
	500  & 9.94  & 4.97  & 4.97  \\
%---------------------------------     
	1000 & 19.9  & 9.95  & 9.95  \\
%--------------------------------- 
\hline\hline
\end{tabular}
%}
\caption{Representative two-body decay widths 
of $\eta$ to $\tau$-flavored leptons,
with $M_{N_1}=5\GeV$ and other $N_k$ decoupled.}
\label{tab:scalar_decays}
\end{table}

In Fig.~\ref{fig:etaC_width} we show the total width of $\eta^\pm$,
defined as 
\begin{align}
\label{eq:width_total_eta}
    \Gamma_{\eta^+}\ &=\ \sum_{\alpha=e}^\tau\ 
    \Gamma(\eta^+\to\ell_\alpha^+ N_k)
\end{align}
as a function of scalar mass $m_\eta$ 
for the representative Majorana neutrino masses 
    $M_{N_1}=50$ ({\color{blue}solid}), 
    $100$ ({\color{Goldenrod}dash}), 
    $150$ ({\color{magenta}dot}), 
    and $200\GeV$ ({\color{teal}dash-dot}),
    with $|Y_{\alpha k}|=1$.
The masses of $N_2$ and $N_3$ are fixed according to 
Eq.~\eqref{eq:scoto_inputs}.

We observe that each curve starts 
at the corresponding two-body threshold
and quickly achieves values 
of $\Gamma_{\eta^+}\sim\mathcal{O}(1)\GeV$. 
Increasing $M_{N_1}$ moves the threshold 
to larger $m_{\eta^\pm}$. 
For $m_{\eta^\pm}\gg M_{N_1}$, the phase
space suppression becomes small and the curves converge.
For $m_{\eta^\pm}\sim 400\GeV\ (1\TeV)$ we find
$\Gamma_{\eta^+}\sim 20\GeV\ (60\GeV)$,
which correspond to lifetimes of $\tau_\eta \sim 10^{-25}s$.
Representative two-body decay widths 
of $\eta$ to $\tau$-flavored leptons 
and assuming $M_{N_1}=5\GeV$
are given in Table~\ref{tab:scalar_decays}.

In the above we neglected the masses of neutrinos and 
neutrino mixing. To reintroduce this and obtain the 
decays of $\eta_R,\eta_I$ to mass eigenstate $\nu_i$,
one makes the following replacements 
in Eq.~\eqref{eq:neutral-width} and 
Eq.~\eqref{eq:neutral-width-other}: 
\begin{subequations}
\label{eq:Ycal}
\begin{align}
    \nu_\alpha\ &\to\ \nu_i 
    \\
    Y_{\alpha k}\ &\to\ 
	\mathcal Y_{ik}
	=
	\left[(U_{\alpha k}^{\rm PMNS})^\dagger Y_{\alpha k}\right]_{ik}
	=
	\sum_\alpha U_{\alpha i}^*Y_{\alpha k}\ .
\end{align}
\end{subequations}
Here, $U_{\alpha k}^{\rm PMNS}$ is the 
PMNS matrix defined in Eq.~\eqref{eq:numixing}.
In this case, one must also sum over both 
$\nu_\alpha$ and $\overline{\nu_\alpha}$ channels
since the mass eigenstates $\nu_i$ are Majorana fermions
in accordance with
$(\mathcal{M}_\nu^{\rm 1-loop})_{\alpha\beta}$
in Eq.~\eqref{eq:neutrino-mass}.

\begin{figure}[!t]
	\centering
	\includegraphics[width=\columnwidth]{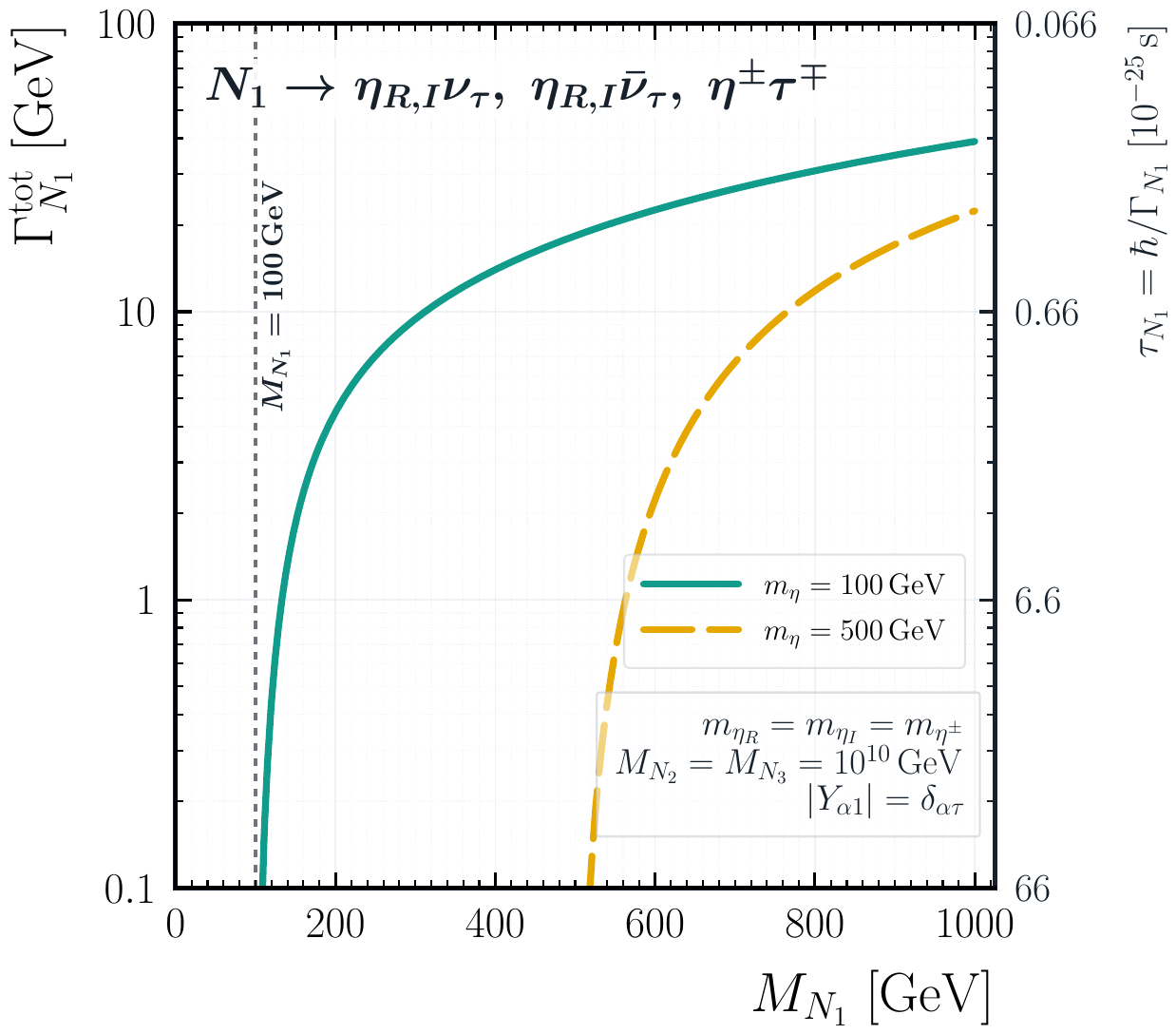}
	\caption{The total width $\Gamma_{N_1}$ [GeV]
    as defined in Eq.~\eqref{eq:width_total_n1}
    as a function of Majorana mass $M_{N_1}$ [GeV] 
    for the scalar masses 
    $m_\eta=100\GeV$ ({\color{teal}solid})
    and 
    $500$ ({\color{Goldenrod}dash}) 
    with $|Y_{\alpha k}|=\delta_{\alpha\tau}$.
	}
	\label{fig:N1_width}
\end{figure}

\begin{table}[t!]
\centering
\resizebox{\columnwidth}{!}{
\begin{tabular}{c c c c c c}
\hline\hline
	$M_{N_1}$
	& $m_\eta$ 
	& $\Gamma_{N_1\to\nu_\tau\eta_R}$
	& $\Gamma_{N_1\to\nu_\tau\eta_I}$
	& $\Gamma_{N_1\to\tau^+\eta^-}$
	& $\Gamma_{N_1}$ [Eq.~\eqref{eq:width_total_n1}]
\\
{[GeV]} & [GeV] & [GeV] & [GeV] & [GeV] & [GeV]
\\
\hline
%--------------------------------- 
    10  & 5     & $28.0\times10^{-3}$  & $28.0\times10^{-3}$  & $54.1\times10^{-3}$  & 0.220    \\
    200 & 100   & 0.560  & 0.560  & 1.12  & 4.48    \\
%--------------------------------- 
    600 & 100   & 2.82  & 2.82  & 5.64  & 22.6  \\
%---------------------------------     
    1000 & 500  & 2.80  & 2.80  & 5.60  & 22.4  \\
%--------------------------------- 
\hline\hline
\end{tabular}
}
\caption{Representative two-body decay widths 
of $N_1$ to $\tau$-flavored leptons,
for various $(m_{N_1},m_\eta)$ configurations.}
\label{tab:neutrino_decays}
\end{table}

%%-----------------------------------------------------------------
%%-----------------------------------------------------------------
\subsection{Majorana Neutrino Decays to Scalars}

Assuming the mass hierarchy, 
$M_{N_k} > m_{\eta^\pm}+m_{\ell_\alpha}$, 
the decay rates for the processes 
\begin{equation}
	N_k\to\ell_\alpha^\pm\eta^\mp\ 
    \quad\text{and}\quad
    N_k\to\nu_\alpha\ \eta_{R,I}
\end{equation}
are given at lowest order by
\begin{align}
\Gamma(N_k\to\ell_\alpha^+\eta^-)\ 
&=\ 
\frac{M_{N_k}}{32\pi}\ |Y_{\alpha k}|^2\
\lambda_K^{1/2}
\left(
1,
\frac{m_{\ell_\alpha}^2}{M_{N_k}^2},
\frac{m_{\eta^\pm}^2}{M_{N_k}^2}
\right)
\nonumber\\
&\qquad\times
\left(
1+
\frac{m_{\ell_\alpha}^2}{M_{N_k}^2}
-\frac{m_{\eta^\pm}^2}{M_{N_k}^2}
\right).
\label{eq:singlet-width-individual}
%-----------------------------------------------
\\
\Gamma(N_k\to\nu_\alpha\eta_R)\
&=\
\frac{M_{N_k}}{64\pi}
|Y_{\alpha k}|^2
\left(
1-\frac{m_{\eta_R}^2}{M_{N_k}^2}
\right)^2.
\label{eq:singlet-neutral-width}
\end{align}

These decay widths were previously reported 
in Refs.~\cite{Hessler:2016kwm,Wang:2026wid}
but under the assumption that $m_{\ell_\alpha}\to0$. 

Under CP symmetry, assuming the appropriate mass hierarchies, 
and with the appropriate exchanges of mass ratios in $\lambda_{K}$, 
we also have the decay widths 
\begin{align}
%------------------------------------
\Gamma(N_k\to\ell_\alpha^-\eta^+)\ &=\ 
    \Gamma(N_k\to\ell_\alpha^+\eta^-)
%------------------------------------
\\    
\Gamma(N_k\to\overline{\nu_\alpha}\eta_R)\ &=\ 
\Gamma(N_k\to\nu_\alpha\eta_I)\ =\ 
\Gamma(N_k\to\overline{\nu_\alpha}\eta_I)\
\nonumber\\
&=\ \Gamma(N_k\to\nu_\alpha\eta_R)\
%------------------------------------
\end{align}

In Fig.~\ref{fig:N1_width} we show the total width 
of $N_1$ as given by 
\begin{align}
\label{eq:width_total_n1}
\Gamma_{N_1}\ =&\ 
\sum_{a=R,I}
\left[\Gamma(N_1\to\nu_\tau\eta_a)
+\ \Gamma(N_1\to\overline{\nu_\tau}\eta_a)\right]
\nonumber\\
&
+\ \Gamma(N_1\to\tau^+\eta^-)\ 
+\ \Gamma(N_1\to\tau^-\eta^+)\ 
\end{align}
as a function of mass $M_{N_1}$ [GeV] 
with 
$m_\eta=100\GeV$ ({\color{teal}solid})
    and 
    $500$ ({\color{Goldenrod}dash}).
Other masses and Yukawa couplings 
are set according Eq.~\eqref{eq:scoto_inputs}.
As with the decays of the $\eta$ fields, 
the two-body decays of $N_1$ open at threshold.
At fixed $M_{N_2}$,
a lighter scalar spectrum leaves more phase
space and gives a larger width. 
Just beyond threshold the total widths stay below 
$\Gamma_{N_1}\sim \mathcal{O}(1)\GeV$,
with lifetimes well beyond the 
$\tau_N\sim\mathcal{O}(10^{-24})s$ level.
Even at $M_N\sim1\TeV$ the widths remain at modest 
values of $\Gamma_{N_1}\sim 20-40\GeV$ 
for the scalar masses under consideration.
Representative two-body decay widths 
of $N_1$ to $\tau$-flavored leptons,
for various $(m_{N_1},m_\eta)$ configurations
are given in Table~\ref{tab:neutrino_decays}.

%%-----------------------------------------------------------------
%%-----------------------------------------------------------------
\subsection{Scalar Decays to Gauge Bosons}

\begin{table}[t!]
\centering
\resizebox{\columnwidth}{!}{
\begin{tabular}{c c c c c}
\hline\hline
$m_{\eta_{\rm parent}}$ & $m_{\eta_{\rm child}}$
& $\Gamma_{\eta^+\to W^+\eta_R}$
& $\Gamma_{\eta^+\to W^+\eta_I}$
& $\Gamma_{\eta_R\to Z\eta_I}$
\\
{[GeV]} & [GeV] & [GeV] & [GeV] & [GeV]
\\
\hline
%--------------------------------- 
100     & 5     & $14.2\times10^{-3}$  & $14.2\times10^{-3}$  & $875\times10^{-6}$  \\
%--------------------------------- 
200     & 100   & 0.217   & 0.217  & $66.2\times10^{-3}$  \\
%--------------------------------- 
500     & 100   & 33.2   & 33.2  & 32.3  \\
%--------------------------------- 
1000    & 500   & 133   & 133  & 131  \\
%--------------------------------- 
%--------------------------------- 
%--------------------------------- 
%--------------------------------- 
\hline\hline
\end{tabular}
}
\caption{Representative two-body decay widths 
for $\eta$ decaying to weak gauge bosons for various $m_\eta$
configurations.}
\label{tab:scalar_decays_gauge}
\end{table}

Assuming the mass hierarchies 
$m_{\eta^\pm}>m_{\eta_R}+M_W$ and $m_{\eta_R} > m_{\eta_I}+M_Z$
the decay rates for the processes
\begin{align}
    \eta^\pm \to W^\pm\ \eta_R\
    \quad\text{and}\quad
    \eta_{R} \to Z\ \eta_I
\end{align}    
are given at lowest order by
\begin{align}
    &\Gamma(\eta^\pm\to W^\pm\eta_R) = 
    \frac{g_W^2}{64\pi} \frac{m_{\eta^\pm}^3}{M_W^2}
    \lambda_K^{3/2}
\left(1,
\frac{M_W^2}{m_{\eta^\pm}^2},
\frac{m_{\eta_R}^2}{m_{\eta^\pm}^2}
\right)
\\
    &\Gamma(\eta_R\to Z \eta_I) = 
    \frac{g_Z^2}{64\pi} \frac{m_{\eta_R}^3}{M_Z^2}
    \lambda_K^{3/2}
\left(1,
\frac{M_Z^2}{m_{\eta_R}^2},
\frac{m_{\eta_I}^2}{m_{\eta_R}^2}
\right) \ .
\end{align}
Taking account of the definitions of $g_Z=g_W/\cos\theta_W$
and $M_Z=M_W/\cos\theta_W$, one finds that the 
two rates are equal, up to the precise values entering
the phase space factor $\lambda_{K}$.
The similarities of these rates reflect 
the underlying SU$(2)_L$ symmetry that the fields 
respect before EW symmetry is broken.

Under CP symmetry, assuming the appropriate mass hierarchies, 
and with the appropriate exchanges of mass ratios in $\lambda_{K}$, 
we also have the decay widths 
\begin{align}
    \Gamma(\eta_1 \to W^\pm\eta_2)\ &=\ \Gamma(\eta^\pm\to W^\pm\eta_R) \ ,
    \\
    \Gamma(\eta_1\to Z \eta_2)\ &=\ \Gamma(\eta_R\to Z \eta_I)\ .
\end{align}
There are no $\eta_1\to h\eta_1$ decays due to energy conservation 
and $\eta_R\to h\eta_I$ decays are forbidden because there is no 
tree-level $\eta_R-\eta_I-h$ vertex.
Representative two-body decay widths 
for $\eta$ decaying to weak gauge bosons for various $m_\eta$
configurations are given in Table~\ref{tab:scalar_decays_gauge}.

%%-----------------------------------------------------------------
%%-----------------------------------------------------------------
\section{Inclusive Production
of Scotogenic Scalars at Hadron Colliders}
\label{sec:production}

\begin{figure*}[!t]
  \centering
  \includegraphics[width=\textwidth]{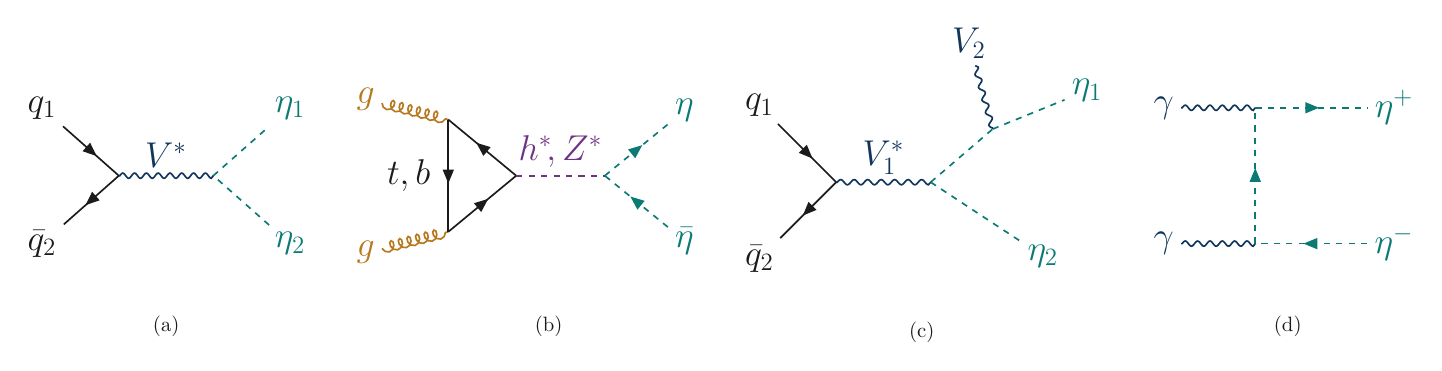}
  \caption{Representative illustrations for the Born-level
  partonic production of Scotogenic scalars pairs $\eta_1,\eta_2$
  in hadron collisions, including
  (a) the Drell-Yan process,
  (b) gluon fusion,
  (c) associated production with a weak boson $V$, and
  (d) photon fusion.
  }
\label{fig:eta_production}
\end{figure*}

In hadron collisions and if kinematically accessible,
the $\eta$ particles can be produced
resonantly through their EW gauge couplings.
Due to the exact $\mathbb{Z}_2$ parity
and fermion number conservation (Lorentz invariance),
the $\eta$ generally must be produced in pairs
or singly in association with an $N_k\ell/N_k\nu_\ell$ lepton pair.

In this section, we present predictions 
for total inclusive cross sections
for the production of $\eta_1\eta_2$ pairs in hadron collisions
up to NLO in QCD with PS matching
via various production mechanisms
at $\sqrt{s}=14\TeV$
(Sec.~\ref{sec:production_xsec}),
at $\sqrt{s}=100\TeV$
(Sec.~\ref{sec:production_xsec_100TeV}).
Differential cross sections are also presented
(Sec.~\ref{sec:production_dxsec}).

To quantify the net impact 
of higher order corrections 
to the Born process,
we define the NLO in QCD $K$-factor ($K^{\rm NLO}$) 
at the cross-section level to be
\begin{align}
 K^{\rm NLO}\ =\ \sigma^{\rm NLO}\ /\ \sigma^{\rm LO}\ .
\end{align}
Similarly, we define the (differential) $K$-factor 
at NLO with respect to observable $\mathcal{O}$ to be
\begin{align}
 K^{\rm NLO}_{\mathcal{O}}\ =\ 
 \frac{d\sigma^{\rm NLO}/d\mathcal{O}}{d\sigma^{\rm LO}/d\mathcal{O}}\ .
\end{align}
We employ the same SM inputs for both 
LO and NLO computations, as specified in Sec.~\ref{sec:setup_sm}.

%%-----------------------------------------------------------------
\subsection{Inclusive Production at 14 TeV}
\label{sec:production_xsec}

\paragraph*{\textbf{Drell-Yan:}}
For scalar masses at or above the EW scale, 
the simplest production modes in hadron collisions 
are the charged-current (CC) and neutral-current (NC)
Drell-Yan channels (DY) at $\mathcal{O}(\alpha^2)$,
\begin{align}
\label{eq:proc_ccdy}
    \text{CCDY}\ &:\ q\overline{q'} \to W^{(*)\pm} 
    \to \eta^\pm \eta^0\ ,
    \\
\label{eq:proc_ncdy}
    \text{NCDY}\ &:\ q\overline{q} \to \gamma^*/Z^{(*)} 
    \to \eta^+\eta^-,\ \eta_R\eta_I\ .
\end{align}
Here and below we denote the neutral scalar and pseudoscalar 
by $\eta^0\ \in\{\eta_{R},\eta_{I}\}$.
Other $\eta_1\eta_2$ configurations at this order 
are forbidden by the $\mathbb{Z}_2$ parity or 
charge conservation.
The partonic process of Eq.~\eqref{eq:proc_ccdy} and Eq.~\eqref{eq:proc_ncdy}
are illustrated diagrammatically 
at the Born level in Fig.~\ref{fig:eta_production}(a)
and have been studied previously at LO in Refs.~\cite{Hessler:2014ssa,Hessler:2016kwm}.

\begin{figure*}[!t]
\subfigure[]{\includegraphics[width=0.45\textwidth]{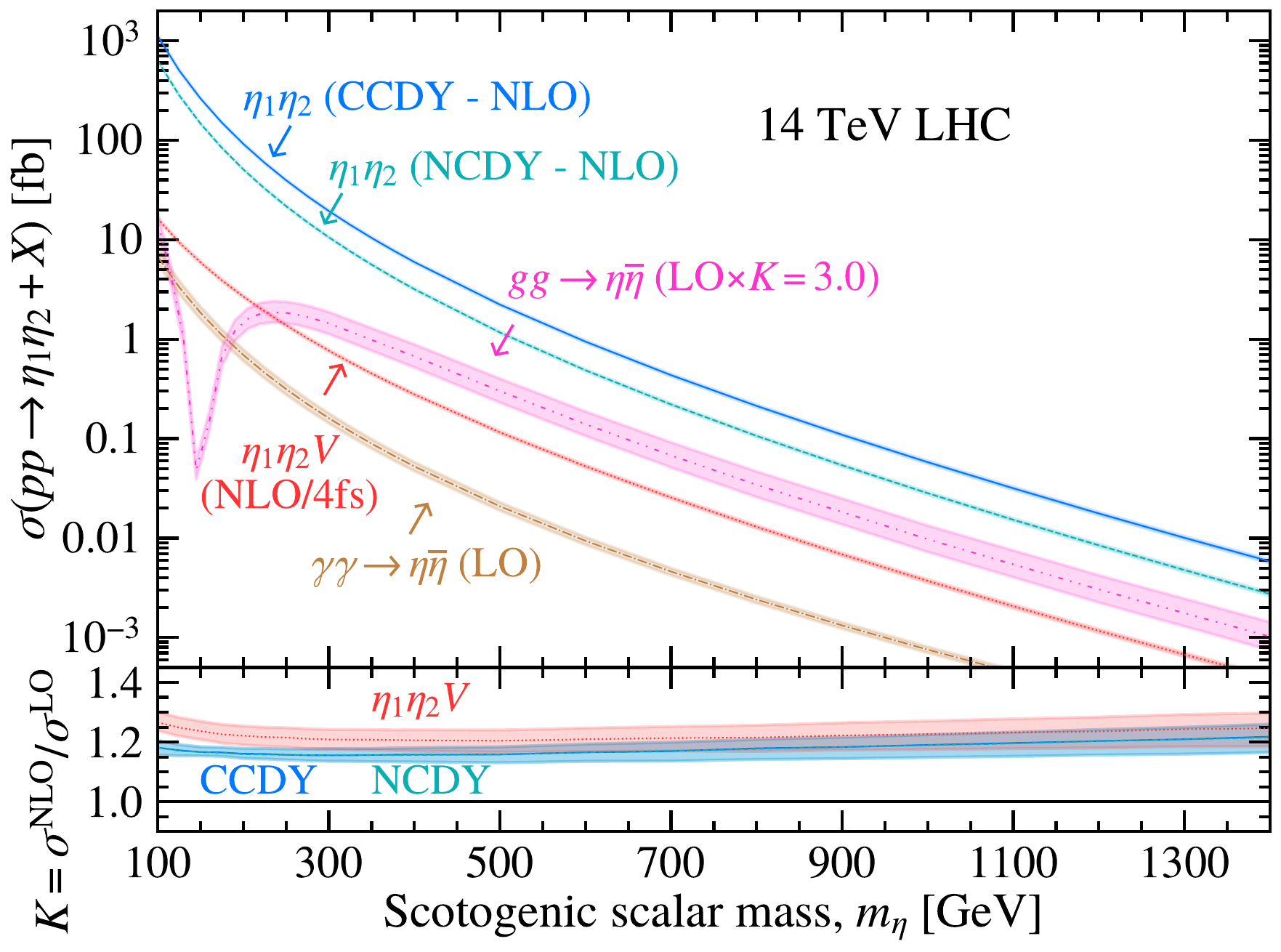}
\label{scotoLHC_xsec_vs_mass_LHCX14}}
\subfigure[]{\includegraphics[width=0.45\textwidth]{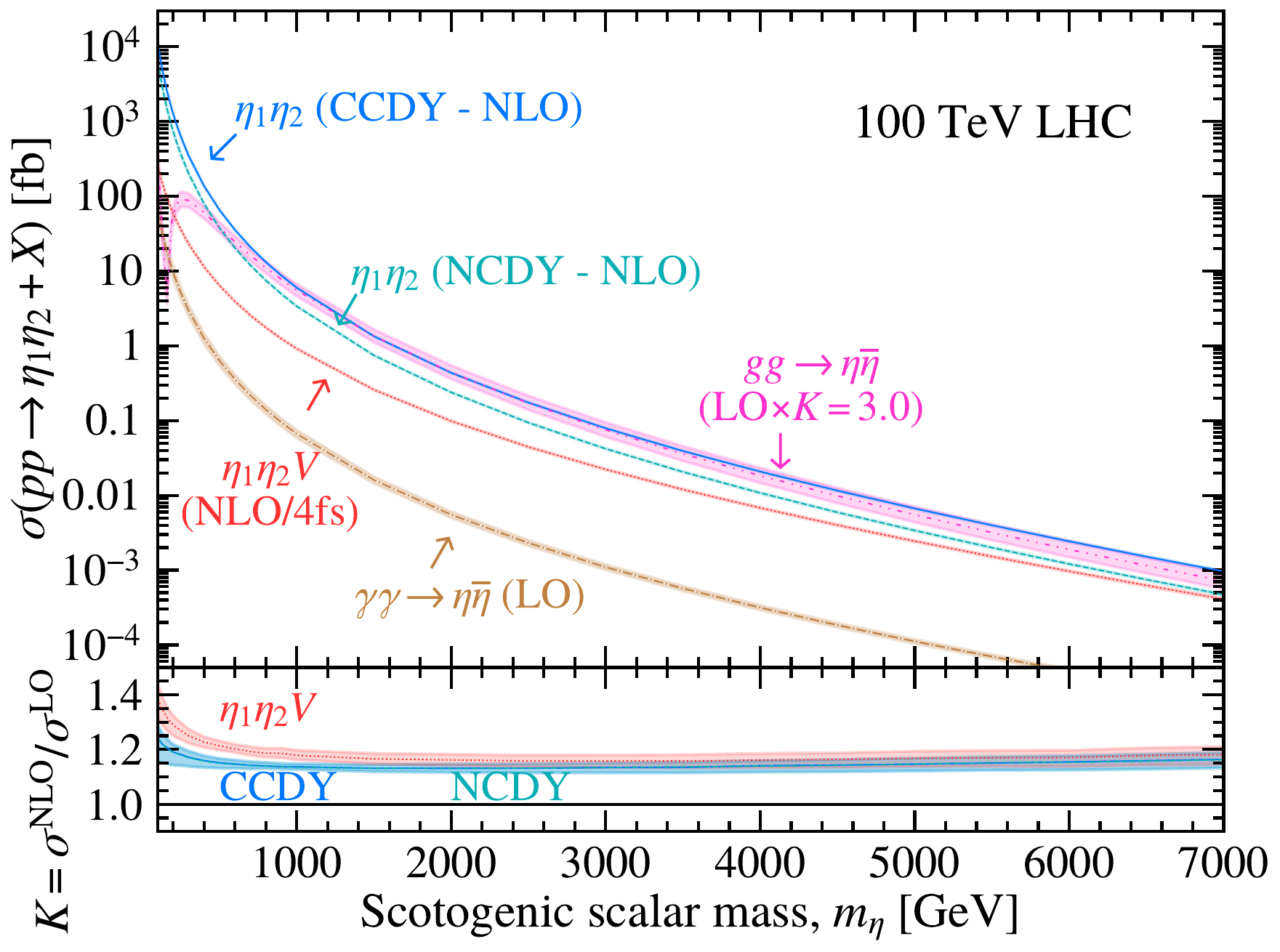}
\label{scotoLHC_xsec_vs_mass_LHC100}}
\caption{Upper: As a function of scalar mass $m_\eta$,
the total inclusive cross section [fb] for the 
CCDY ({\color{blue}solid}),
NCDY ({\color{teal}dash}),
GF ({\color{magenta}dash-dot-dot}),
AF ({\color{brown}dash-dot}), and
$\eta\eta V$ ({\color{red}dot}) channels 
at various perturbative accuracies
with their residual scale 9-point uncertainty band
for the (a) $\sqrt{s}=14\TeV$ LHC
and a hypothetical (b) $\sqrt{s}=100\TeV$ $pp$ collider.
Lower: For the CCDY, NCDY, and $\eta\eta V$ channels,
the NLO in QCD $K$-factor with residual scale uncertainty band.}
\label{fig:scotoLHC_xsec_vs_mass}
\end{figure*}

In Fig.~\ref{scotoLHC_xsec_vs_mass_LHCX14}, 
we show at $\sqrt{s}=14\TeV$
and as a function of the scalar mass
the NLO-accurate cross sections for the
CCDY ({\color{blue}solid})
and
NCDY ({\color{teal}dash})
channels.
For scalar masses in the range
\confirm{$m_{\eta}=100\GeV-1400\GeV$},
we find that the inclusive CCDY cross section at NLO 
$(\sigma_{\rm DY}^{\rm NLO})$,
residual scale and PDF uncertainties 
at NLO $(\delta\sigma_{\rm DY}^{\rm NLO})$,
and the NLO $K$-factors ($K^{\rm NLO}$), 
respectively span
\begin{subequations}
\begin{align}
\label{sec:xsec_LHCX14_CCDY}
	\sigma^{\rm NLO}_{\rm CCDY}\ &=\ \confirm{1.1\times10^{+03}\fb\ -\ 5.8\times10^{-03}\fb}\ ,\\
	\delta\sigma_{\rm CCDY}^{\rm NLO}\ &\sim\ \confirm{^{+3.5\%}_{-4.3\%}\ ({\rm scale})\quad ^{+6\%}_{-6\%}\ ({\rm PDF})}\ ,\\
	K_{\rm CCDY}^{\rm NLO}\ &=\ \confirm{1.16\ -\ 1.22}\ .
\end{align}
\end{subequations}
For the NCDY case, we have the similar values
\begin{subequations}
\begin{align}
\label{sec:xsec_LHCX14_NCDY}
	\sigma^{\rm NLO}_{\rm NCDY}\ &=\ \confirm{6.3\times10^{+02}\fb\ -\ 2.7\times10^{-03}\fb}\ ,\\
	\delta\sigma_{\rm NCDY}^{\rm NLO}\ &\sim\ \confirm{^{+3.5\%}_{-4.3\%}\ ({\rm scale})\quad ^{+6\%}_{-6\%}\ ({\rm PDF})}\ ,\\
	K_{\rm NCDY}^{\rm NLO}\ &=\ \confirm{1.16\ -\ 1.22}\ .
\end{align}
\end{subequations}
As the CC and NC channels both proceed 
by the annihilation of massless quark-antiquark pairs 
into a colorless vector boson at LO,
the structure of their virtual and real radiative corrections 
are identical~\cite{Altarelli:1979ub} 
(see also, e.g., App.~A of Ref.~\cite{Ruiz:2015zca}).
This results is numerically similar values, 
up to permutations of quark flavor.

For the range of scalar masses under consideration, 
the CCDY channel remains consistently above the NCDY channel,
with the ratio at NLO spanning,
\begin{align}
\label{sec:xsec_dy_ratio_LHCX14}
	\sigma^{\rm NLO}_{\rm CCDY}\ /\ \sigma^{\rm NLO}_{\rm NCDY}\ &=\ \confirm{1.77\ -\ 2.12}\ .
\end{align}
The smaller (larger) values of the ratio correspond 
to lower (larger) $\eta$ masses.
These values can be understood 
from the various gauge charges and couplings 
entering the expressions for the partonic DY cross sections.

To illustrate this, we first note that QCD corrections at NLO 
for a generic, high-mass DY process are driven by the finite part 
of a factorizable virtual correction (see, e.g., Ref.~\cite{Ruiz:2015zca}).
Most of the contribution to the inclusive cross section 
from the real radiative channel 
is already included in the normalization of evolved PDFs.
This means that the ratio of NLO cross sections is well approximated by 
the ratio of the cross sections at LO,
\begin{align}
\label{sec:xsec_dy_ratio_theory}
    \frac{\sigma^{\rm NLO}_{\rm CCDY}}{\sigma^{\rm NLO}_{\rm NCDY}}\ 
    \approx\ 
    \frac{f_i\ \otimes\ f_j\ \otimes\ \hat{\sigma}_{ij\to \eta^\pm\eta_R,\  \eta^\pm\eta_I}}
    {f_i\ \otimes\ f_j\ \otimes\ \hat{\sigma}_{ij\to \eta^+\eta^-,\ \eta_R\eta_I}}\ ,
\end{align}
where the sum over parton species $i,j$ is implicit.

The partonic cross sections for the CCDY processes are flavor universal.
Their expressions are given by
\begin{align}
\label{eq:xsec_parton_ccdy}
    \hat{\sigma}_{\rm CCDY}\
    &=\
    \hat{\sigma}_{ij\to \eta^\pm\eta_R}\
    =\ 
    \hat{\sigma}_{ij\to \eta^\pm\eta_I}
    \\
    &= \frac{g_W^4}{3\cdot 2^8 \cdot \pi\ \cdot N_c}\
    \frac{\hat{s}\ \lambda_K^{3/2}(1,r_1,r_2)}{\vert D_W(\hat{s})\vert^2}\ .
\end{align}
Here, $\lambda_K(x,y,z)$ is the K\"allen function defined 
in Eq.~\eqref{eq:kallen} with $r_i = m_{\eta i}^2/\hat{s}$.
$\hat{s}=(p_i+p_j)^2$ is the squared partonic center-of-mass energy.
$g_W\approx0.65$ is the weak coupling constant, 
$N_c=3$ is the number of colors, 
and $D_W(\hat{s})$ is Breit-Wigner pole structure 
for virtuality $\sqrt{\hat{s}}$,
\begin{align}
    D_V(\hat{s})\ &=\ \hat{s}-M_V^2 + i M_V\Gamma_V\ .
\end{align}

The partonic cross sections for the NCDY processes 
depend on the weak isospin $(T_L^3)$ and electromagnetic $(Q_X)$ charges 
of external particles.
For generic charges $(T_L^3)_X$ and $Q_X$, 
the cross sections can be written as~\cite{Gunion:1996pq}
\begin{subequations}
\label{eq:xsec_parton_ncdy}
\begin{align}
%-----------------------------------------------------
 \hat{\sigma}_{\rm NCDY} &= 
 \mathcal{F}\times
 \left[P_{\gamma\gamma} + P_{\gamma Z} + P_{ZZ} \right]\ ,\quad
 \text{where}
%----------------------------------------------------- 
\\
 \mathcal{F} &=\ 
 \frac{\pi\ \alpha_{\rm EM}^2}{2\cdot 3 \cdot N_c}\ 
 \lambda_{K}^{3/2}(1,r_1,r_1)\ , r_i = \frac{m_{\eta i}^2}{\hat{s}}\ ,
 \\
 P_{\gamma\gamma} &=\ \frac{2 Q_f^2 Q_\eta^2}{\hat{s}},
 \\
 P_{\gamma Z} &= 
 \frac{2 Q_f Q_\eta A_\eta\ \left(g_L^f + g_R^f\right)}{s_W^2\ c_W^2}
 \frac{(\hat{s}-M_Z^2)}{\vert D_Z(\hat{s})\vert^2}
 , 
 \\
 P_{ZZ} &= \frac{A^2_\eta\ \left((g_L^f)^2 + (g_R^f)^2 \right)}{s_W^4\ c_W^4}
 \frac{\hat{s}}{\vert D_Z(\hat{s})\vert^2}\ .
 \end{align}
Here, $P_{\gamma\gamma}$ and $ P_{ZZ}$ are the photon and $Z$ 
contributions, $P_{\gamma Z}$ is the $\gamma/Z$ interference,
and coupling factors are
\begin{align}
% \\
 A_\eta &= (T_L^3)_\eta - Q_\eta\ s_W^2\ ,\ s_W = \sin\theta_W\ ,
 \\
 g_L^f &= (T_L^3)_f - Q_f\ s_W^2\ ,\ c_W=\cos\theta_W\ ,
 \\
 g_R^f &= - Q_f\sin^2\theta_W\  .
\end{align}
\end{subequations}

In Eq.~\eqref{eq:xsec_parton_ncdy}, 
the masses of the two outgoing scalars are assumed equal,
as indicated by the velocity factor $\lambda_{K}^{3/2}(1,r_1,r_1)$
in the prefactor $\mathcal{F}$.
For the charged scalars $\eta^\pm$,
this is dictated by electromagnetic current conservation,
i.e., 
$(p_{\eta^+}+p_{\eta^-})_\mu \cdot J^\mu(p_{\eta^+},p_{\eta^-})=0$.
For the neutral scalars $\eta_R$, $\eta_I$, 
the mass splitting set by the $H$-$\eta$ doublet coupling $\lambda_5$ 
as given in Eq.~\eqref{eq:neutral-splitting}.
The corresponding NCDY cross section 
for $m_{\eta_R}\neq m_{\eta_I}$ is 
\begin{align}
\label{eq:xsec_parton_ncdy_alt}
    \hat{\sigma}_{ij\to \eta_R\eta_I}
    &= \frac{g_Z^4\ \left((g_L^f)^2 + (g_R^f)^2 \right)}{3\cdot 2^7 \cdot \pi\ \cdot N_c}
    \frac{\hat{s}\ \lambda_K^{3/2}(1,r_1,r_2)}{\vert D_Z(\hat{s})\vert^2}\ ,
\end{align}
where $g_Z=g_W/c_W$.
Setting $r_2=r_1$ recovers Eq.~\eqref{eq:xsec_parton_ncdy} 
for the gauge quantum numbers $((T_L^3)_X,Q_X)=(\frac{1}{2},0)$.

In the ``low'' $m_\eta$ regime, 
the DY currents at $\sqrt{s}=14\TeV$
are dominated by sea-quark-sea-antiquark annihilation,
with all parton species contributing comparably to the process.
For parton PDF $f_i(x,\mu_f)$ 
at fixed momentum fraction $x$ and scale $\mu_f$,
$f_{u}\approx f_{\overline{u}}\approx f_d\approx \dots$
This means that the impact of PDFs 
in Eq.~\eqref{sec:xsec_dy_ratio_theory} cancel,
 giving
\begin{align}
\label{sec:xsec_dy_ratio_lo_mass}
    \frac{\sigma^{\rm NLO}_{\rm CCDY}}{\sigma^{\rm NLO}_{\rm NCDY}}\ 
    &\Big\vert_{\rm low\ mass}
    \nonumber\\
    &\approx\ 
    \frac{4\ \hat{\sigma}_{\rm CCDY}}{
    \sum_{i=u.d}\ 
    (\hat{\sigma}_{i\overline{i}\to \eta^+\eta^-}+
    \hat{\sigma}_{i\overline{i}\to \eta_R\eta_I})
    }
    \\
    &= \frac{18\ c_W^4}{9-18 s_W^2 + 20 s_W^4}\ 
    \approx\ 1.80\ .
\end{align}

This value is in good agreement with the lower end 
of Eq.~\eqref{sec:xsec_dy_ratio_LHCX14}.
To reach the final line,
we neglected the masses of $W$ and $Z$ appearing 
in the Breit-Wigner factor $D_V(\hat{s})$.
The factor of $4$ appearing in the numerator 
of Eq.~\eqref{sec:xsec_dy_ratio_lo_mass}
is due to summing over 
the four channels $\eta^\pm \eta_R^0$, $\eta^\pm \eta_I^0$.
A sum over the number of quark generations is assumed to cancel 
between the numerator and 
denominator\footnote{\label{foot:charmPDF}In
reality, the $s\overline{s}$ subchannel survives and increases 
the denominator while the $s\overline{c}$, $\overline{s}c$, 
and $c\overline{c}$ subchannels are suppressed due to the smallness
of the $c$,$\overline{c}$ PDFs.
Correctly modeling this would likely decrease the estimate 
in Eq.~\eqref{sec:xsec_dy_ratio_lo_mass} and improve agreement.}.

In the ``high'' $m_\eta$ regime, DY production is mediated 
predominantly by valence-quark-sea-quark annihilation.
In this regime, the $u$-quark PDF is enhanced 
over the $d$-quark PDF, 
which we parameterize as $f_u \approx (1+\epsilon_u)f_d$.
The region is also characterized by a modest $\overline{u}-\overline{d}$
asymmetry, with $f_{\overline{d}}$ being 
larger~\cite{NNPDF:2021njg,NNPDF:2024djq,FNALE906SeaQuest:2025kjo}. 
We parameterize this asymmetry 
as $f_{\overline{d}}\approx (1+\epsilon_{\overline{d}})f_{\overline{u}}$.
With these modifications,
the ratio of NLO cross sections is approximately
\begin{align}
\label{sec:xsec_dy_ratio_hi_mass}
    \frac{\sigma^{\rm NLO}_{\rm CCDY}}{\sigma^{\rm NLO}_{\rm NCDY}}\ 
    &\Big\vert_{\rm high\ mass}
    \nonumber\\
    \approx\ &
    \frac{2\ [(1+\epsilon_u)(1+\epsilon_{\overline{d}}) + 1]\ \hat{\sigma}_{\rm CCDY}}{
    \sum_{i=u.\overline{d}}\ (1+\epsilon_i)
    (\hat{\sigma}_{i\overline{i}\to \eta^+\eta-}+
    \hat{\sigma}_{i\overline{i}\to \eta_R\eta_I})
    }
    \\
    =\ & \frac{18\ c_W^4\ (\tilde{\epsilon}+\epsilon_u\epsilon_{\overline{d}})}{9\tilde{\epsilon}-18 s_W^2\tilde{\epsilon} 
    + 2 s_W^4 (20+13\epsilon_u + 7\epsilon_{\overline{d}})}\ ,
    \\
    \tilde{\epsilon} \equiv\ & (2+\epsilon_u + \epsilon_{\overline{d}})\ .
\end{align}

The factor of 2 in the numerator of Eq.~\eqref{sec:xsec_dy_ratio_hi_mass}
stems from summing over $\eta_R^0$ and $\eta_I^0$.
The factor of 1 in the bracket is the $d\overline{u}$ contribution 
while the $(1+\epsilon_u)(1+\epsilon_{\overline{d}})$ factor is the 
``enhanced'' $u\overline{d}$ factor.
The $(1+\epsilon_i)$ factor in the denominator 
similarly accounts for differences in PDFs.
We again neglect the masses of $W$ and $Z$ appearing 
in the Breit-Wigner factor $D_V(\hat{s})$.

To estimate the value of $\epsilon_u$ and $\epsilon_{\overline{d}}$
we evaluate directly the PDFs in \texttt{python3}
using the \texttt{LHAPDF6} interface
at the values 
$x=2m_\eta/\sqrt{s}$ and $\mu_f=m_\eta$.
Explicitly, we do
\begin{verbatim}
>>> import lhapdf
>>> lhaid = 335900
>>> nnPDF = lhapdf.mkPDF(lhaid)
>>> mH = 1400
>>> epsU = nnPDF.xfxQ(2,mH/7000.,mH)/
            nnPDF.xfxQ(1,mH/7000.,mH) - 1.0
>>> epsDX = nnPDF.xfxQ(-1,mH/7000.,mH)/
            nnPDF.xfxQ(-2,mH/7000.,mH) - 1.0
\end{verbatim}
For $m_\eta = 1400\GeV$ and $\sqrt{s}=14\TeV$, we obtain 
\begin{align}
    \epsilon_u\ \approx\  1.18 \quad\text{and}\quad 
    \epsilon_{\overline{d}}\ \approx\ 0.47\ .
\end{align}
Inserting these values into Eq.~\eqref{sec:xsec_dy_ratio_hi_mass}
gives a ratio of
\begin{align}
    \frac{\sigma^{\rm NLO}_{\rm CCDY}}{\sigma^{\rm NLO}_{\rm NCDY}}\ 
    \Big\vert_{\rm high\ mass}\ 
    \approx\ 2.04\ ,
\end{align}
which is within reasonable agreement \confirm{(within $4\%$)}
with the upper end 
of the ratio in Eq.~\eqref{sec:xsec_dy_ratio_LHCX14}.

If charged scalar pairs are discovered at the LHC,
cross section ratios will be paramount to 
discriminating among competing hypotheses,
particularly for radiative neutrino mass 
models~\cite{delAguila:2013yaa,Cai:2017jrq,Ruiz:2022sct}.
The exercise of estimating Eq.~\eqref{sec:xsec_dy_ratio_LHCX14}
demonstrates a clear role of sea-quark distributions,
and their relative differences, 
in the DY production channels.
Knowledge of gauge charges and gauge couplings 
was insufficient to reproduce the full range,
and reinforces the importance 
of having a precise knowledge of PDFs
in searches for new physics.

%%-----------------------------------------------------------------
\paragraph*{\textbf{Gluon Fusion:}}

Beyond the DY channels are the 
loop-induced gluon fusion (GF) channels 
at $\mathcal{O}(\alpha^2 \alpha_s^2)$,
\begin{align}
   \text{GF}\ &:\  gg\ \to\ \eta^+\eta^-,\ \eta_R\eta_R,\ \eta_I\eta_I\ .
\end{align}
In the context of the Scotogenic model, 
these channels were previously studied in Ref.~\cite{Hessler:2014ssa}.

At lowest order, the GF channel proceeds through 
a heavy-quark loop and intermediate $h^{*}/Z^*$ exchanges,
as illustrated in Fig.~\ref{fig:eta_production}(b).
Formally, however, 
the channel is a finite, separately gauge-invariant contribution
to the inclusive NCDY channel 
at next-to-next-to-leading order in QCD.
Importantly, proceeding through the Higgs means that the channel 
is sensitive to trilinear $h-\eta-\overline{\eta}$ couplings,
$\lambda_3$, $\lambda_L$, and $\lambda_S$, as given in 
Eq.~\eqref{eq:higgs-portal}.

Like other channels of the form of Fig.~\ref{fig:eta_production}(b),
the GF channel in the Scotogenic model is sensitive to large QCD corrections.
To approximate these, we use the $K$-factor
\begin{align}
    K^{\rm N3LL}\ =\ \sigma^{\rm N3LL}\ /\ \sigma^{\rm LO}\ \approx 3.0\ ,
\end{align}
which is based on next-to-next-to-next-to-leading log (N3LL) 
threshold resummed predictions for particles produced via GF 
in other Seesaws models over similar mass 
ranges~\cite{Ruiz:2017yyf,Cai:2017mow,Pascoli:2018heg,Fuks:2019clu}.
We do not propagate / reduce the residual uncertainties 
due to missing real corrections~\cite{Ruiz:2017yyf}.

The cross section for the GF channel ({\color{magenta}dash-dot-dot})
is shown in Fig.~\ref{fig:scotoLHC_xsec_vs_mass}.
For the same scalar masses as before
the scattering rates and 
uncertainties span
\begin{subequations}
\begin{align}
\label{sec:xsec_LHCX14_GGF}
	\sigma^{\rm LO}_{\rm GGF}\times K^{\rm N3LL}\ &=\ \confirm{45\fb\ -\ 3.1\ab}\ ,\\
	\delta\sigma_{\rm GGF}^{\rm LO}\ &\sim\ \confirm{^{+40\%}_{-27\%}\ ({\rm scale})\quad ^{+4\%}_{-4\%}\ ({\rm PDF})}\ .
\end{align}
\end{subequations}

The QCD-corrected GF rate at 14 TeV sits 
\confirm{one-to-two orders of magnitude}
below the NCDY current over the whole mass range.
We caution that this ratio is subject to the 
trilinear coupling, including its sign, due 
to interference with the $Z^*$ diagram~\cite{Hessler:2014ssa}.
The ``dip'' at $m_\eta\approx 150\GeV$ 
should be interpreted as an enhancement at $m_\eta\approx 170\GeV$
and is due to the momentum 
of the internal top quarks surpassing threshold $(2m_\eta \gtrsim 2m_t)$.

%%-----------------------------------------------------------------
\paragraph*{\textbf{Associated Production}}
In addition to the pair production channels above, 
one can also consider the associated production channels 
$\eta_1\eta_2 B$ or even $\eta_1\eta_2 B_1 B_2$, 
where $B$ is an EW boson $B\in\{W^\pm,Z,\gamma,h\}$.
Such multiboson channels are sensitive 
to the size and sign of scalar couplings 
but a complete investigation is left to future work.

For concreteness and to minimize the dependence on scalar couplings,
we consider the $\mathcal{O}(\alpha^3)$ channel,
\begin{align}
\eta\eta V\ &:\ q\overline{q'}, q\overline{q}\ \to\ \eta_1\eta_2 V,
\end{align}
where $V\in\{W^\pm,Z\}$, 
as illustrated diagrammatically in Fig.~\ref{fig:eta_production}(c).
Importantly, the same-sign scalar channels $\eta^\pm\eta^\pm W^\mp$
are forbidden by the $\mathbb{Z}_2$ parity\footnote{For the same reason, 
the same-sign $WW$ scattering channel $W^\pm W^\pm \to \eta^\pm\eta^\pm$ 
is also forbidden.}.
In addition, to avoid nuances related 
intermediate top quarks that appear 
at NLO in $b$-initiated channels, 
we restrict ourselves 
to the $n_f=4$ active quark flavor scheme\footnote{i.e., using the \texttt{SM\_Scotogenic\_MassiveLeptons\_4fs\_NLO} UFO.}.

The predicted cross sections for the $\eta\eta V$ channel ({\color{red}dot}) 
are shown in Fig.~\ref{fig:scotoLHC_xsec_vs_mass}.
For scalar masses under consideration, 
we find that 
the predicted cross sections of the $\eta\eta V$ channel at NLO,
the residual scale and PDF uncertainties at NLO,
and the NLO $K$-factors all span
\begin{subequations}
\begin{align}
\label{sec:xsec_LHCX14_HHV}
	\sigma^{\rm NLO}_{\eta_1\eta_2V}\ &=\ \confirm{16\fb\ -\ 0.4\ab}\ ,\\
	\delta\sigma_{\eta_1\eta_2V}^{\rm NLO}\ &\sim\ \confirm{^{+4.1\%}_{-4.9\%}\ ({\rm scale})\quad ^{+8\%}_{-8\%}\ ({\rm PDF})}\ ,\\
	K_{\eta_1\eta_2V}^{\rm NLO}\ &=\ \confirm{1.20\ -\ 1.27}\ .
\end{align}
\end{subequations}

Just beyond the top quark threshold, 
the cross sections for $\eta\eta V$ sit just below the GF channel.
The scaling with $m_\eta$ mirrors the DY channels because both 
sets of channels are initiated at LO by quark-antiquark annihilation.
QCD corrections at NLO are slightly larger than the DY channels due 
to the diboson configurations.
As reported in Ref.~\cite{Binoth:2008kt},  
$q\overline{q}\to V_1 V_2^*$ configurations exhibit slightly larger 
virtual corrections at $\mathcal{O}(\alpha_s)$
due to additional box diagrams 
that are absent in pure DY topologies.

\begin{table*}[t!]
\begin{center}
\resizebox{\textwidth}{!}{
\begin{tabular}{c | c | r r c || r r c}
\hline\hline
\multicolumn{2}{c}{} & \multicolumn{3}{c}{$\sqrt{s} = 14\TeV$ LHC} & \multicolumn{3}{c}{$\sqrt{s} = 100\TeV$ LHC}
\\
mass [GeV] & Process & $\sigma^{\rm LO}\ [\rm fb]\ \delta_{\rm scale}\ \delta_{\rm PDF}$ & $\sigma^{\rm NLO}\ [\rm fb]\ \delta_{\rm scale}\ \delta_{\rm PDF}$ & $K$ & $\sigma^{\rm LO}\ [\rm fb]\ \delta_{\rm scale}\ \delta_{\rm PDF}$ & $\sigma^{\rm NLO}\ [\rm fb]\ \delta_{\rm scale}\ \delta_{\rm PDF}$ & $K$
\\
\hline\hline
\multirow{5}{*}{$250$} & CCDY & ${3.42 \cdot 10^{1}}^{+3.0\%}_{-3.0\%}\ ^{+1.2\%}_{-1.2\%}$ & ${3.97 \cdot 10^{1}}^{+1.7\%}_{-1.5\%}\ ^{+1.2\%}_{-1.2\%}$ & $1.16$ & ${5.24 \cdot 10^{2}}^{+6.1\%}_{-6.9\%}\ ^{+0.4\%}_{-0.4\%}$ & ${6.21 \cdot 10^{2}}^{+2.2\%}_{-3.3\%}\ ^{+0.4\%}_{-0.4\%}$ & $1.19$ \\
 & NCDY & ${1.89 \cdot 10^{1}}^{+2.9\%}_{-2.9\%}\ ^{+1.0\%}_{-1.0\%}$ & ${2.18 \cdot 10^{1}}^{+1.7\%}_{-1.5\%}\ ^{+1.0\%}_{-1.0\%}$ & $1.15$ & ${3.07 \cdot 10^{2}}^{+6.4\%}_{-7.1\%}\ ^{+0.4\%}_{-0.4\%}$ & ${3.60 \cdot 10^{2}}^{+2.0\%}_{-3.2\%}\ ^{+0.4\%}_{-0.4\%}$ & $1.17$ \\
 & GF & $1.84^{+29.4\%}_{-21.4\%}\ ^{+0.8\%}_{-0.8\%}$ &  &  & ${9.09 \cdot 10^{1}}^{+25.9\%}_{-20.6\%}\ ^{+0.7\%}_{-0.7\%}$ &  &  \\
 & $\eta\eta V$ & $1.13^{+5.6\%}_{-5.1\%}\ ^{+1.4\%}_{-1.4\%}$ & $1.38^{+2.7\%}_{-2.6\%}\ ^{+1.4\%}_{-1.4\%}$ & $1.22$ & ${2.79 \cdot 10^{1}}^{+3.3\%}_{-3.9\%}\ ^{+0.4\%}_{-0.4\%}$ & ${3.54 \cdot 10^{1}}^{+2.3\%}_{-2.6\%}\ ^{+0.4\%}_{-0.4\%}$ & $1.27$ \\
 & AF & $0.31^{+12.6\%}_{-12.1\%}\ ^{+1.3\%}_{-1.3\%}$ &  &  & $5.03^{+19.4\%}_{-17.4\%}\ ^{+0.9\%}_{-0.9\%}$ &  &  \\
\hline
\multirow{5}{*}{$1000$} & CCDY & ${4.87 \cdot 10^{-2}}^{+11.2\%}_{-9.5\%}\ ^{+3.6\%}_{-3.6\%}$ & ${5.78 \cdot 10^{-2}}^{+2.9\%}_{-3.5\%}\ ^{+3.6\%}_{-3.6\%}$ & $1.19$ & $5.24^{+0.5\%}_{-0.8\%}\ ^{+0.8\%}_{-0.8\%}$ & $5.95^{+1.0\%}_{-0.8\%}\ ^{+0.8\%}_{-0.8\%}$ & $1.14$ \\
 & NCDY & ${2.36 \cdot 10^{-2}}^{+11.1\%}_{-9.5\%}\ ^{+3.4\%}_{-3.4\%}$ & ${2.83 \cdot 10^{-2}}^{+2.9\%}_{-3.5\%}\ ^{+3.4\%}_{-3.4\%}$ & $1.20$ & $2.94^{+0.4\%}_{-0.7\%}\ ^{+0.7\%}_{-0.7\%}$ & $3.36^{+1.1\%}_{-0.8\%}\ ^{+0.7\%}_{-0.7\%}$ & $1.14$ \\
 & GF & ${0.97 \cdot 10^{-2}}^{+37.5\%}_{-25.6\%}\ ^{+2.7\%}_{-2.7\%}$ &  &  & $5.50^{+21.1\%}_{-16.6\%}\ ^{+0.6\%}_{-0.6\%}$ &  &  \\
 & $\eta\eta V$ & ${3.02 \cdot 10^{-3}}^{+12.5\%}_{-10.5\%}\ ^{+4.8\%}_{-4.8\%}$ & ${3.71 \cdot 10^{-3}}^{+3.6\%}_{-4.2\%}\ ^{+4.7\%}_{-4.7\%}$ & $1.23$ & $0.77^{+2.0\%}_{-2.0\%}\ ^{+0.9\%}_{-0.9\%}$ & $0.91^{+1.6\%}_{-1.4\%}\ ^{+0.9\%}_{-0.9\%}$ & $1.18$ \\
 & AF & ${7.55 \cdot 10^{-4}}^{+7.0\%}_{-7.2\%}\ ^{+2.3\%}_{-2.3\%}$ &  &  & ${6.63 \cdot 10^{-2}}^{+12.3\%}_{-11.7\%}\ ^{+1.0\%}_{-1.0\%}$ &  &  \\
\hline
\hline
\end{tabular}}
\caption{Total cross section for the CCDY, NCDY, GF, $\eta \eta V$ and AF channels for $m_{\eta} = 250$ GeV and $m_{\eta} = 1$ TeV, at $\sqrt{s} = 14$ TeV and $\sqrt{s} = 100$ TeV. The table also contains the scale uncertainties, PDF uncertainties, and QCD K-factor.}
\label{tab:xsec-table}
\end{center}
\end{table*}

%%-----------------------------------------------------------------
\paragraph*{\textbf{Photon Fusion}}
Finally, we have pair production of charged scalars 
at $\mathcal{O}(\alpha^4)$ via photon fusion (AF),
\begin{align}
   \text{AF}\ &:\  \gamma\gamma\ \to\ \eta^+\ \eta^-\ ,
\end{align}
which is illustrated diagrammatically in Fig.~\ref{fig:eta_production}(d).
The channel proceeds entirely through electromagnetic gauge charges, 
and therefore is insensitive at LO to the weak isospin gauge charges.

The cross sections for the AF channel ({\color{brown}dash-dot})
are shown in Fig.~\ref{fig:scotoLHC_xsec_vs_mass}.
For the scalar masses under consideration 
the scattering rates and uncertainties span
\begin{subequations}
\begin{align}
\label{sec:xsec_LHCX14_AAF}
	\sigma^{\rm LO}_{\rm AAF}\ &=\ \confirm{6.8\fb\ -\ 0.1\ab}\ ,\\
	\delta\sigma_{\rm AAF}^{\rm LO}\ &\sim\ \confirm{^{+17.4\%}_{-16.0\%}\ ({\rm scale})\quad ^{+3\%}_{-3\%}\ ({\rm PDF})}\ .
\end{align}
\end{subequations}

In comparison to the CCDY channel, the AF rates are about 
\confirm{one-to-two orders of magnitude} smaller.
The AF channel at LO also carries a somewhat modest $\mathcal{O}(\pm15\%)$
scale uncertainty, which can be reduced 
by matching to channels with higher leg multiplicities, 
namely the $q\gamma \to q \eta^+\eta^-$ 
and $qq'\to qq' \eta^+\eta^-$ channels
at $\mathcal{O}(\alpha^3)$ and $\mathcal{O}(\alpha^4)$.

%%-----------------------------------------------------------------
\subsection{Inclusive Production at 100 TeV}
\label{sec:production_xsec_100TeV}

Turning our focus to higher collider energies,
in Fig.~\ref{scotoLHC_xsec_vs_mass_LHC100}
we show the production cross section for the same channels 
as in Sec.~\ref{sec:production_xsec} 
but for $\sqrt{s}=100\TeV$.

For the mass range $m_\eta = 100\GeV-7\TeV$,
NLO-accurate cross sections span roughly
\begin{subequations}
\begin{align}
\label{sec:xsec_LHC100}
	\sigma^{\rm NLO}_{\rm CCDY}\ &=\ \confirm{11\pb\ -\ 0.96\ab}\ ,\\
    \sigma^{\rm NLO}_{\rm NCDY}\ &=\ \confirm{6.4\pb\ -\ 0.47\ab}\  ,\\
	\sigma^{\rm LO}_{\rm GGF}\times K^{\rm N3LL}\ &=\ \confirm{1.0\pb\ -\ 2.1\ab}\ ,\\
	\sigma^{\rm NLO}_{\eta_1\eta_2V}\ &=\ \confirm{250\fb\ -\ 0.41\ab}\ ,\\
	\sigma^{\rm LO}_{\rm AAF}\ &=\ \confirm{63\fb\ -\ 0.02\ab}\ .
\end{align}
\end{subequations}

For quark-initiated channels and for the lowest $m_\eta$ under consideration, 
this is about a \confirm{10-fold increase} 
compared to scattering rates at $\sqrt{s}=14\TeV$.
For the GF channel,
this is about a \confirm{20-fold increase} 
due to the much faster growth in gluon PDFs at low $x$.

For the DY channels, the scale uncertainties 
exhibit small $\mathcal{O}(1\%-2\%)$ increases 
while the PDF uncertainties feature a comparable reduction.
The first originates from the increasing importance of 
quark-gluon and antiquark-gluon partonic subprocesses,
but which are only described at LO when the inclusive process 
is described at NLO.
The second originates from probing PDFs at lower momentum fractions
at 100 TeV than at 14 TeV,
\begin{align}
 \max&(x_{100\TeV}) \approx \frac{(2\times7\TeV)}{(100\TeV)}\ 
 \approx\ 0.14
 \nonumber\\
 &<\ 
 \max(x_{14\TeV}) \approx \frac{(2\times1.4\TeV)}{(14\TeV)}\ 
 \approx\ 0.2\ ,
\end{align}
where PDFs are generally better modeled by modern fits.

For the other channels 
similar reductions in the PDF uncertainties are observed.
The GF and $\eta\eta V$ both feature comparable scale uncertainties.
Finally, the AF channel features a slight increase in scale uncertainty.

For the DY and $\eta\eta V$ channels, the QCD $K$-factors at 100 TeV
are comparable to their 14 TeV counter parts, 
except for at ``low'' masses $(m_\eta \lesssim 500\GeV)$.
For lower masses $K$-factors are slightly larger at 100 TeV
due to the jump in gluon density at low $x$ and the increasing importance 
of $qg$ and $\overline{q}g$ partonic subchannels.

The ratio of the CC and NCDY channels spans
\begin{subequations}
\begin{align}
\label{sec:xsec_dy_ratio_LHC100}
	\sigma^{\rm NLO}_{\rm CCDY}\ /\ \sigma^{\rm NLO}_{\rm NCDY}\ &=\ \confirm{1.71\ -\ 2.04}\ .
\end{align}
\end{subequations}
This remains consistent with the ratio at $\sqrt{s}=14\TeV$
in Eq.~\eqref{sec:xsec_dy_ratio_LHCX14},
and the expectations following the arguments given in the previous 
section (see also footnote \ref{foot:charmPDF}).

For representative $m_\eta$, cross sections at LO and NLO, 
uncertainties, and $K$-factors
are listed in Table~\ref{tab:xsec-table}.

\begin{figure*}[!t]
\subfigure[]{\includegraphics[width=0.45\textwidth]{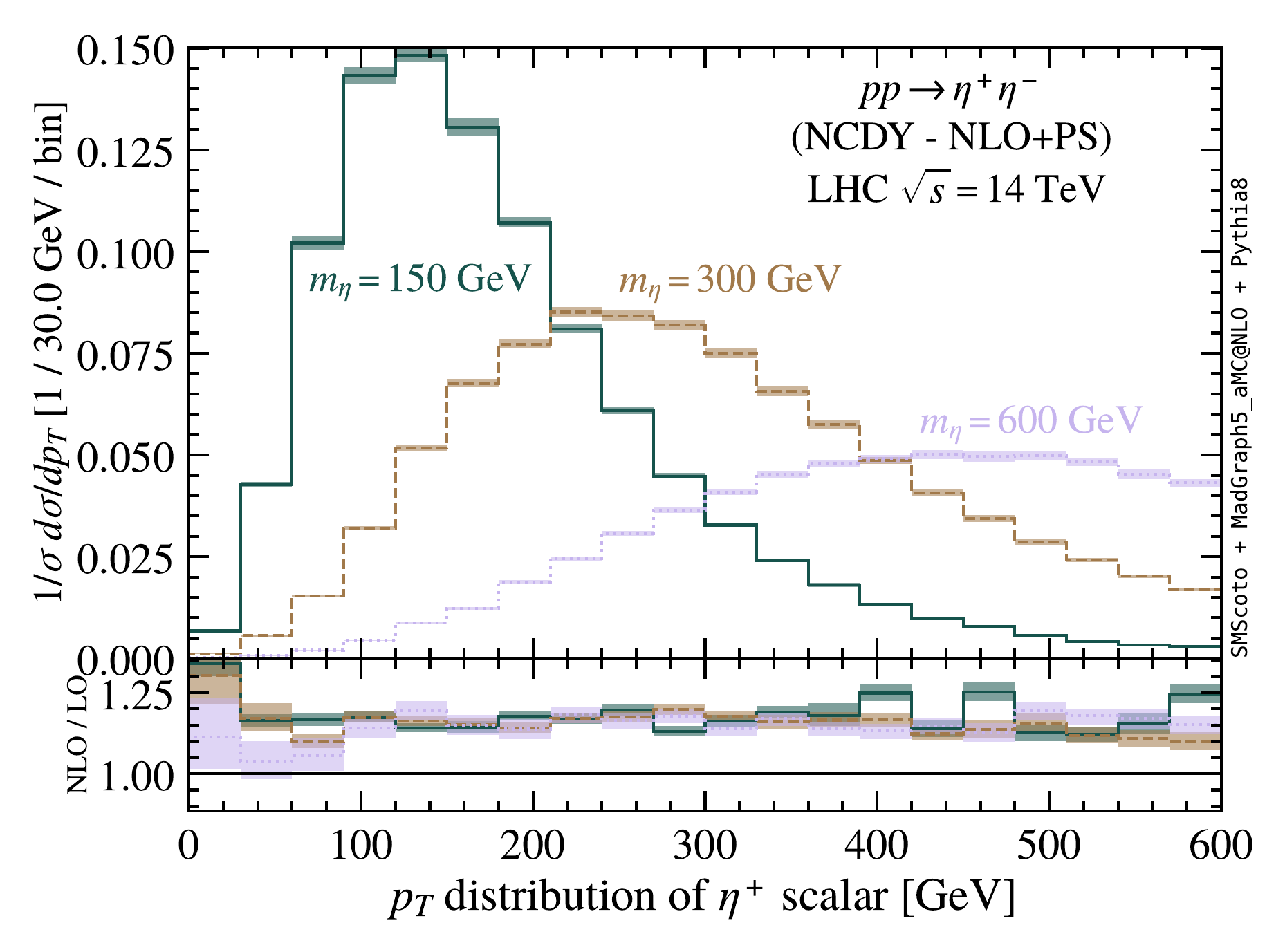}
\label{fig:scotoLHC_NCDY_HpHm_NLOPS_pT_LHCX14_MultiMass}}
\subfigure[]{\includegraphics[width=0.45\textwidth]{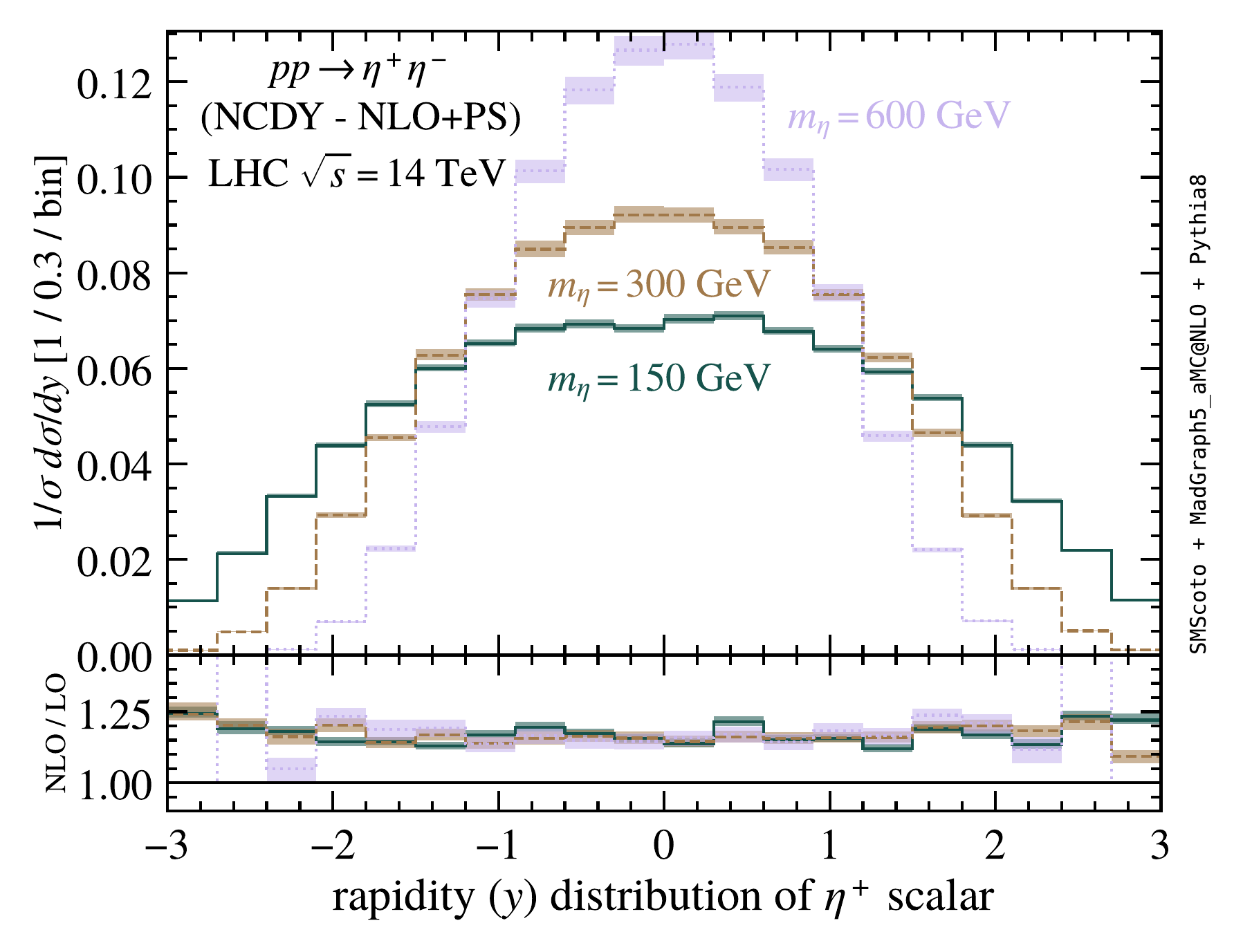}
\label{fig:scotoLHC_NCDY_HpHm_NLOPS_yX_LHCX14_MultiMass}}
\\
\subfigure[]{\includegraphics[width=0.45\textwidth]{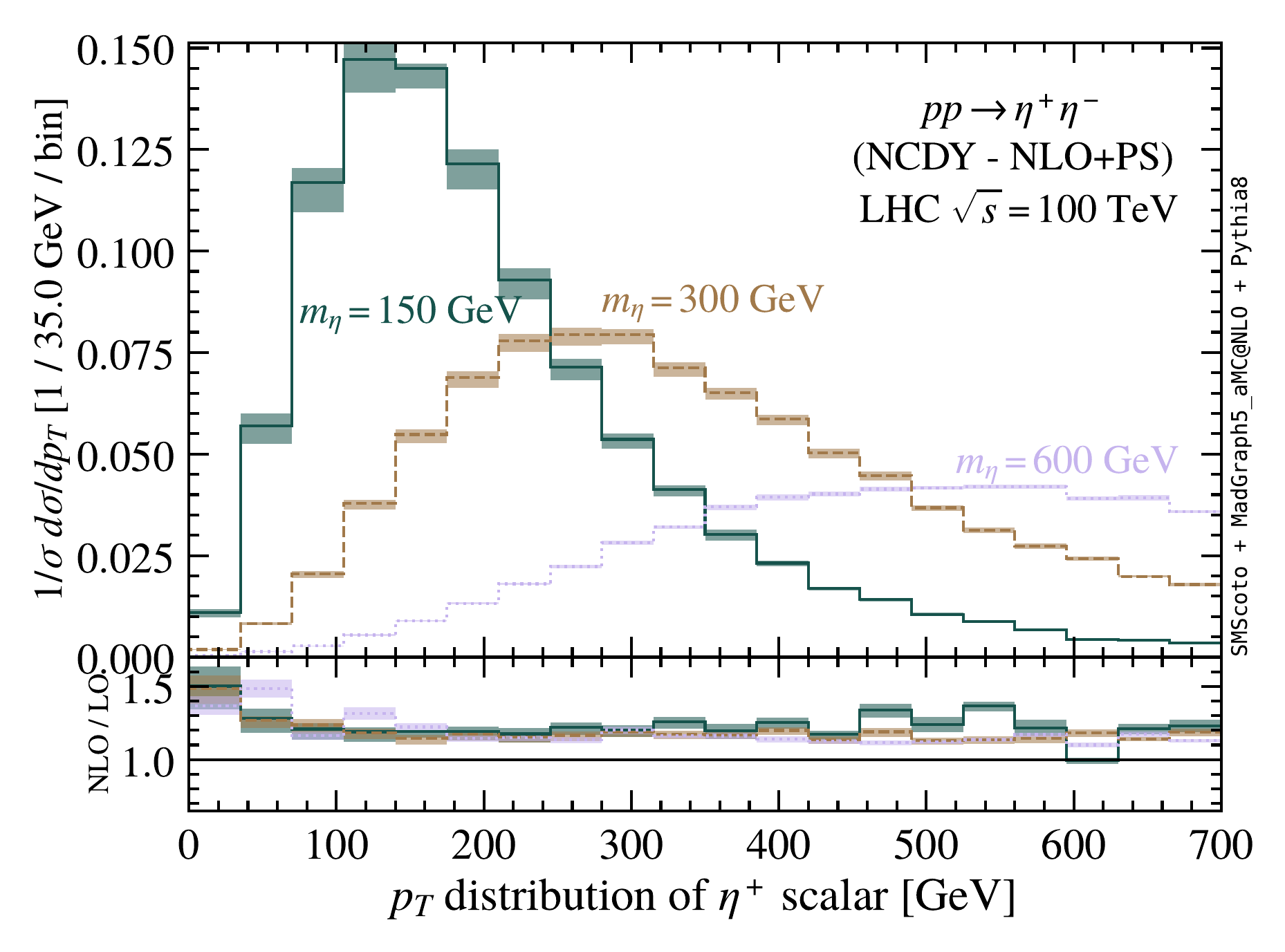}
\label{fig:scotoLHC_NCDY_HpHm_NLOPS_pT_LHC100_MultiMass}}
\subfigure[]{\includegraphics[width=0.45\textwidth]{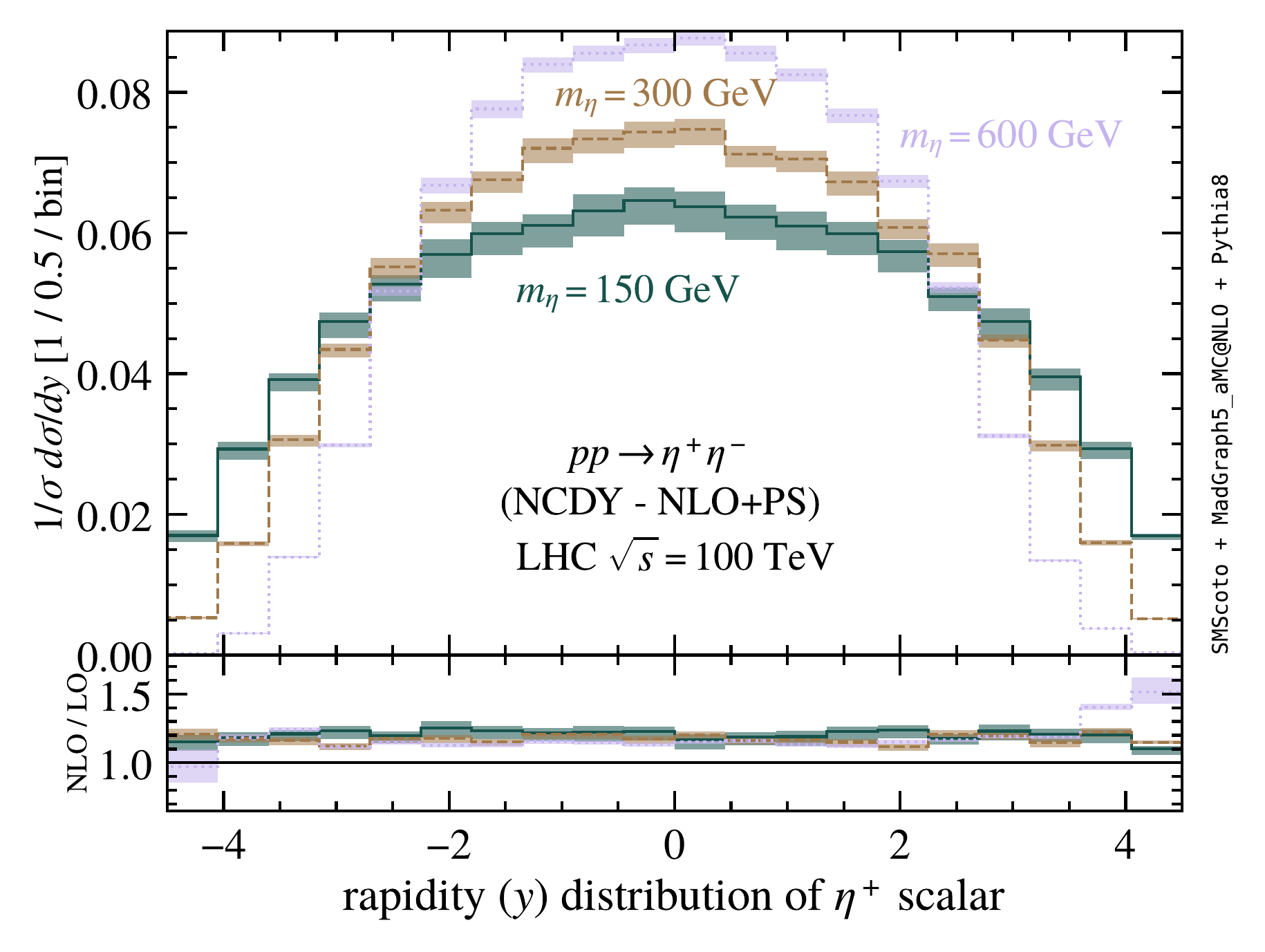}
\label{fig:scotoLHC_NCDY_HpHm_NLOPS_yX_LHC100_MultiMass}}
\caption{Upper: The normalized differential distributions 
$(1/\sigma\cdot d\sigma/d\hat{\mathcal{O}})$ 
with respect to (a,c) transverse momentum $(p_T)$ and (b,d) rapidity $(y)$ 
of the scalar $\eta^+$ in the NCDY process at NLO+PS 
with 9-point scale uncertainties 
for (a,b) $\sqrt{s}=14\TeV$ and (c,d) $\sqrt{s}=100\TeV$,
and representative masses
$m_\eta = 150\GeV$ ({\color{darkgreen}solid}),
$450\GeV$ ({\color{brown}dash}),
and
$600\GeV$ ({\color{magenta}dotted}).
Lower: Ratio with respect to the distribution at LO+PS.}
\label{fig:scotoLHC_NCDY_HpHm_NLOPS}
\end{figure*}

%%-----------------------------------------------------------------
\subsection{Differential Cross Sections}
\label{sec:production_dxsec}

Hadron collisions at the TeV scale 
can be incredibly rich and complex due 
to the interplay between 
hard-scattering dynamics and soft subprocesses.
This interplay is driven by a number of variables
and manifests in differential distributions 
and fiducial cross sections.
Among these variables are the absolute scale of the hard-scattering process,
$Q\sim \mathcal{O}(m_\eta)$,
and the relative scale at which the incoming hadrons are probed, 
$\xi \sim Q / \sqrt{s} \sim \mathcal{O}(m_\eta/\sqrt{s})$.
Naturally, these are tied to the mass of the particles being produced 
and to the collider energy itself, among other factors.

To briefly explore this interplay 
in the Scotogenic model, 
we show in Fig.~\ref{fig:scotoLHC_NCDY_HpHm_NLOPS_pT_LHCX14_MultiMass}
and Fig.~\ref{fig:scotoLHC_NCDY_HpHm_NLOPS_yX_LHCX14_MultiMass}
the normalized distributions $(1/\sigma\cdot d\sigma/d\hat{\mathcal{O}})$ 
with respect to the (a,c) transverse momentum $(\hat{\mathcal{O}}=p_T)$ 
and (b,d) rapidity $(\hat{\mathcal{O}}=y)$,
\begin{align}
    p_T &=\ \sqrt{p_x^2\ +\ p_y^2}\ =\ \vert \vec{p}^{\rm lab}\vert \sin\theta^{\rm lab}\,\
    \\
    y &= 
    \frac{1}{2} \log\left(\frac{E^{\rm lab}+p_z^{\rm lab}}{E^{\rm lab}-p_z^{\rm lab}}\right) 
    = 
    \log\left(\frac{E^{\rm lab}+p_z^{\rm lab}}{\sqrt{m^2 + p_T^2}}\right)\ ,
\end{align}
of the scalar $\eta^+$ in the NCDY process
\begin{align}
    p p\ \to\ \eta^+\ \eta^-\ 
\end{align}
at NLO in QCD with parton shower (PS) matching
at (a,b) $\sqrt{s}=14\TeV$ and (c,d) $\sqrt{s}=100\TeV$ 
for the representative masses
$m_\eta = 150\GeV$ ({\color{darkgreen}solid}),
$450\GeV$ ({\color{brown}dash}),
and
$600\GeV$ ({\color{magenta}dotted}).
Also shown are the 9-point scale uncertainty bands and  
the ratio with the distribution at LO+PS (lower panels).

Focusing first on the $p_T$ distributions (a,c), we observe at both 
collider configurations the distributions plateau 
at $p_T\sim (4/5)\times m_\eta$,
with endpoints at $p_T=0$ and power-law suppression $d\sigma\sim1/p_T^k$ 
(with $k>0$) at larger $p_T$.
This behavior can be attributed to the behavior of the 
partonic cross section, as given Eq.~\eqref{eq:xsec_parton_ncdy}.

At partonic threshold $\hat{s}=4m_\eta^2$, 
the $\eta^\pm$ scalars have little-to-no kinetic energy
since $\beta = \sqrt{1-4 m_\eta^2/\hat{s}}$,
and hence little-to-no transverse momentum.
This configuration corresponds to the endpoint.
At large scattering scales, all mass scales can be neglected 
and the partonic DY cross sections scale as 
$\hat{\sigma}_{\rm DY} \sim 1/\hat{s}\sim 1/\vert \vec{p}\vert^2\sim 1/p_T^2$. 
This configuration corresponds to the high-$p_T$ tail.

For the plateau, we follow the arguments 
of Ref.~\cite{Pascoli:2018heg,Cirigliano:2021peb}
and first note that the invariant mass distributions 
for DY-type processes at LO scale as
\begin{align}
\label{eq:xsec_mass_scaling}
    \frac{d\sigma_{\rm DY}^{\rm LO}}{d\sqrt{\hat{s}}}\ =\
    \left(\frac{\sqrt{\hat{s}}}{s}\right)\ 
    \Phi(\hat{s})\ \otimes\
    \hat{\sigma}_{\rm DY}(m_\eta^2,\hat{s})\ .
\end{align}
Here, $\sqrt{s}$ and $\sqrt{\hat{s}}=\sqrt{x_1x_2 s}$ 
are the collider and partonic 
center-of-mass energies,
and 
$\Phi(\hat{s})$ is the partonic luminosity
that is convolved $(\otimes)$
with the $\hat{\sigma}_{\rm DY}^{\rm LO}$
given in Eqs.~\eqref{eq:xsec_parton_ccdy} and \ref{eq:xsec_parton_ncdy}.
We assume that the DY channel is dominated 
by low-$x$ partons, which implies the scaling 
\begin{align}
\label{eq:luminosity_low_x}
    \Phi(\hat{s})\ \sim\ 
    f_q(x_1)f_{\overline{q}'}(x_2)\
    \overset{x_i\to0}\sim\ 
    \frac{1}{x_1}\frac{1}{x_2}\ 
    =\ \frac{s}{\hat{s}}\ .
\end{align}

Taking the derivative of Eq.~\eqref{eq:xsec_mass_scaling} 
with respect to $\sqrt{\hat{s}}$
and setting the result to zero gives three extrema: 
(i) $\sqrt{\hat{s}}\to 2m_\eta$, 
(ii) $\sqrt{\hat{s}}\to \infty$,
and 
(iii) $\sqrt{\hat{s}}\to 2\sqrt{2}m_\eta$.
Configuration (i) corresponds to threshold production;
(ii) corresponds to a minimum in the tail 
of invariant mass distribution of the $(\eta^+\eta^-)$-system
at asymptotic energies;
and 
(iii) corresponds to a global maximum.
Finally, for a fixed $\sqrt{\hat{s}}$, 
explicit calculation for the \textit{average} $p_T$ 
for either final-state scalar 
at the partonic level is given by 
\begin{align}
\langle p_T\rangle\ &=\ 
\frac{1}{\sigma_{\rm DY}}\int_{-1}^{+1} d\cos\theta\ \cdot\ \frac{d\hat{\sigma}}{d\cos\theta}\ \cdot\ p_T
\\
&=\ \frac{9\pi}{64}\ \sqrt{\hat{s}}\ \sqrt{1-\frac{4m_\eta^2}{\hat{s}}}\ .
\end{align}
The expression is exact and holds for both CCDY and NCDY 
under the assumption that $m_{\eta_{R,I}} = m_{\eta^\pm}$.

Since $p_T$ is invariant under boosts along the beam axis,
up to real radiative corrections,
we can estimate the average $p_T$ of $\eta^\pm$
for the plateau region in the lab frame 
using the extrema values of Eq.~\eqref{eq:xsec_mass_scaling}.
Doing this, we obtain
\begin{align}
    \confirm{\langle p_T\rangle_{\rm plateau}\ 
    =\ \frac{9\pi}{32} \times m_\eta\ 
    \approx\ 0.88\times m_\eta\ },
\end{align}
which is in reasonable agreement with the NLO+PS distributions 
at $\sqrt{s}=14\TeV$ and $\sqrt{s}=100\TeV$.
The exception is $m_\eta=600\GeV$ at $\sqrt{s}=14\TeV$,
which plateaus at lower $p_T$ due 
to additional phase space suppression 
in Eq.~\eqref{eq:luminosity_low_x}
as $4m_\eta^2\to s$.

Focusing now on the $y$ distributions (b,d), 
we observe a migration of events 
from smaller absolute rapidity $\vert y\vert $ 
to larger $\vert y\vert$ as $(m_\eta/ \sqrt{s})\to0$.
At $\sqrt{s}=14\TeV$, we find that rapidities reach upwards 
of $\vert y\vert\sim2-3$ for $m_\eta=150-600\GeV$ 
while at $\sqrt{s}=100\TeV$ we find that rapidities reach 
upwards of $\vert y\vert\sim4$ for the same masses.

This behavior reflects three general trends:
(i) Lighter (heavier) objects are produced with more (less) kinetic energy 
for a fixed collision scale $\hat{s}=x_1x_2s \geq 4m_\eta^2$, 
where $x_i= 2E_i/\sqrt{s}$ are the momentum fractions carried 
by the incoming partons. 
This leads to heavier objects being produced with smaller rapidities.

(ii) The threshold condition $x_1x_2= (\hat{s}/s) \geq 4m_\eta^2/s$
can be satisfied with more (fewer) asymmetric $(x_1,x_2)$ configurations 
at larger (smaller) $\sqrt{s}$.
This means that the momentum of the $(\eta^+\eta^-)$-system 
in the lab frame (at LO), 
$P_{\rm sys}^\mu=(\sqrt{s}/2)(x_1+x_2,0,0,x_1-x_2)$, can carry 
larger $z$ momentum $P_{\rm sys}^{\mu=3}=\sqrt{s}(x_1-x_2)$
at $\sqrt{s}=100\TeV$, 
which then propagates into the $z$ momentum of each $\eta^\pm$.

We find that QCD corrections at $\mathcal{O}(\alpha_s)$ 
impact kinematical observables at LO+PS
by about $\mathcal{O}(+25\%)$.
Differential $K$-factors $K_\mathcal{O}^{\rm NLO}$
are largely flat over the ranges under investigation.
This is consistent with other studies of Seesaw particles 
and scalars produced in DY channels~\cite{Cai:2017mow}.
We caution, however, that these modest corrections 
also reflect the modeling prescriptions in Sec.~\ref{sec:setup}.

\begin{figure*}[!t]
    \centering
\subfigure[]{\includegraphics[width=0.45\textwidth]{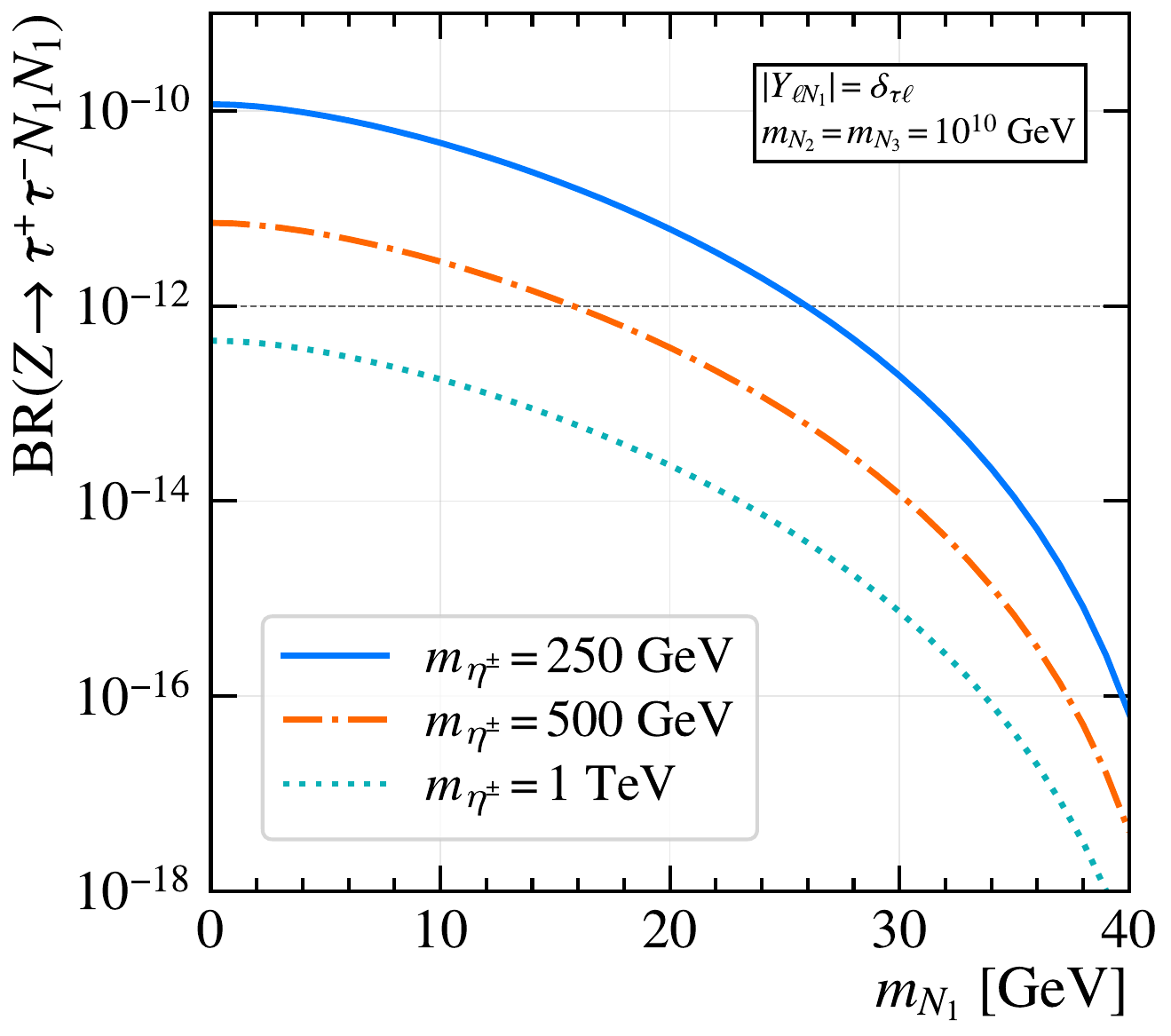}}
\subfigure[]{\includegraphics[width=0.45\textwidth]{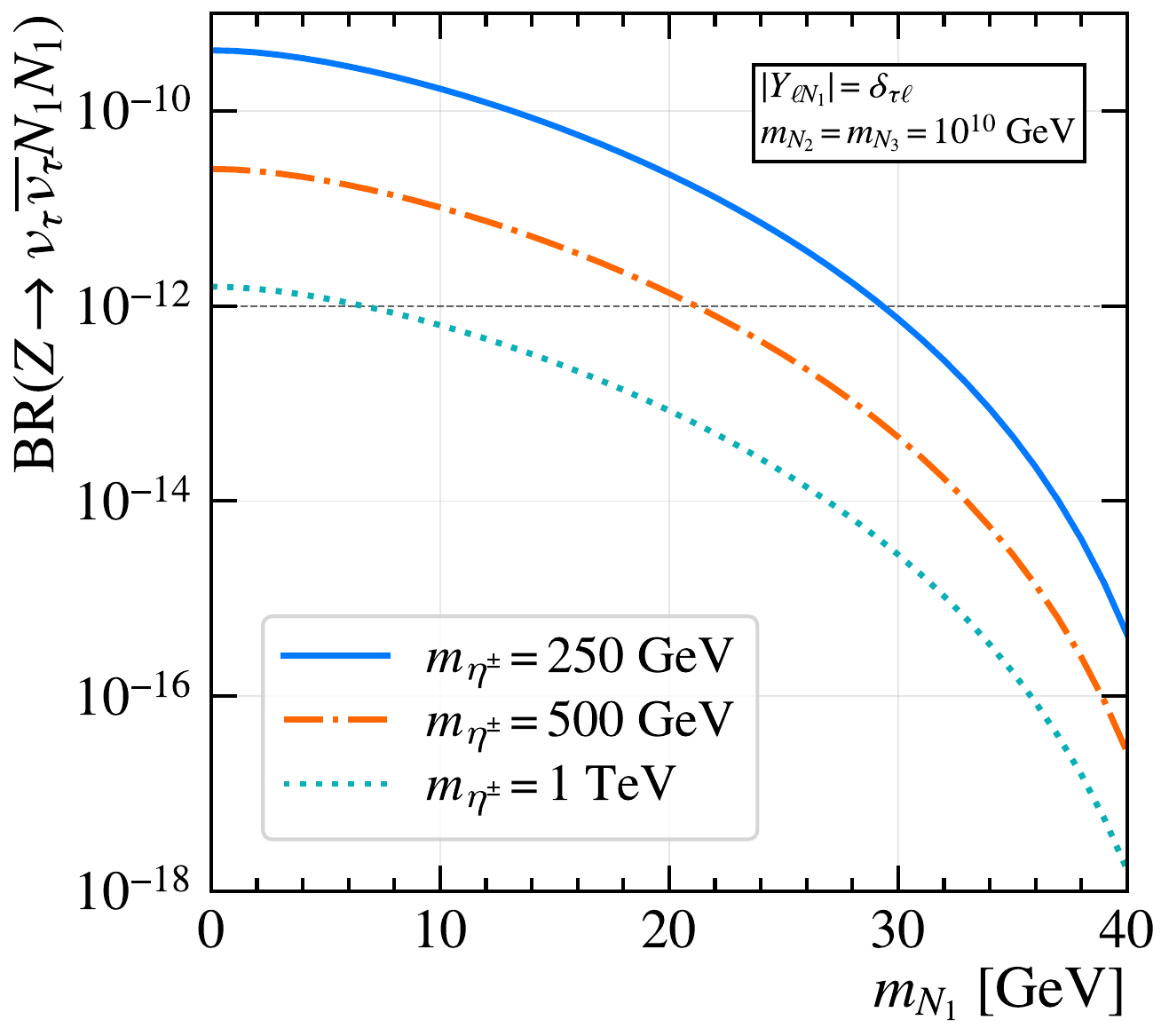}}
\caption{Branching rate for the decay processes 
(a) $Z \to \tau^{+}\tau^{-}N_{1}N_{1}$ 
and 
(b) $Z \to \nu_{\tau} \overline{\nu_{\tau}}N_{1}N_{1}$,
as a function of $M_{N_1}$ and 
assuming $m_{\eta^{\pm}} = 250$ GeV ({\color{blue}solid}), 
$m_{\eta^{\pm}} = 500$ GeV ({\color{orange}dot-dashed}), and 
$m_{\eta^{\pm}} = 1$ TeV ({\color{teal}dotted}), 
while $m_{\eta^\pm}=m_{\eta_R}=m_{\eta_I}$.}
\label{fig:BRsZ_rare}
\end{figure*}

%%-----------------------------------------------------------------
%%-----------------------------------------------------------------
\section{Outlook and Future Colliders}
\label{sec:outlook}

%%-----------------------------------------------------------------
\subsection{Rare Z Decays at Z Factories}
\label{sec:fccee}

Beyond the production in many-TeV hadron collisions,
Scotogenic particles can also be produced in $e^+e^-$ collisions,
particularly through decays of the $W$ and $Z$ boson.
For example: assuming the $N_k$ are sufficiently light, 
then new, ultra rare decay channels of the $W$ and $Z$ include
\begin{subequations}
\begin{align}
    W^\pm\ \to\ & \ell^\pm_\alpha \nu_\beta N_k N_{k'}\ ,\\
    Z\ \to\ & \ell^-_\alpha\ell^+_\beta N_k N_{k'}\ ,\\
    Z\ \to\ & \nu_\alpha\overline{\nu_\beta} N_k N_{k'}\ ,
\end{align}
\end{subequations}
which are mediated by far-off-shell $\eta^\pm$ and $\eta_{R/I}$.
For lepton flavors $\alpha\neq\beta$, these decay modes 
also violate the conservation of lepton flavor.
At the $Z$ pole run of the FCC-ee, $\sim\mathcal{O}(10^{12})$ $Z$ bosons
will be produced with \confirm{$\mathcal{L} \simeq 150\invab$ of data}~\cite{FCC:2018evy}.
This means that decay rates as low as ${\rm BR}(Z\to X)\sim 10^{-11}-10^{-9}$
can be probed.

As an illustration of the physics potential at the FCC-ee, 
we show in Fig.~\ref{fig:BRsZ_rare} 
the $Z\to \tau^+\tau^-N_1 N_1$ 
and 
the $Z\to \overline{\nu_\tau}\nu_\tau N_1 N_1$ 
branching rates as a function of $M_{N_1}$
for representative mass and coupling configuration
\begin{align}
m_{\eta^\pm}=m_{\eta_R}&=m_{\eta_I} = 250\GeV\ (500\GeV)\ [1\TeV]
\nonumber\\
m_{N_2} &= m_{N_3} = 10^{10}\GeV\ ,\ Y_{\ell N_{k}}=\delta_{\tau \ell}\ .
\end{align}
We sum over all interfering diagrams but consider only the $\tau$-flavor 
channel; the mass of $\tau$ is kept as in Eq.~\eqref{eq:sm_inputs}.

As it can be seen from Fig.~\ref{fig:BRsZ_rare}, considering the available phase space for a real $Z$ boson to decay into a pair of $N_{1}$s 
and two SM leptons, i.e., 0 $< M_{N_1} \lesssim$ 40 GeV, the FCC-ee could, in principle, probe these channels, 
especially for lower values of $m_{\eta},M_{N_k}$. 
The $Z \to \nu_{\tau} \overline{\nu_{\tau}}N_{1}N_{1}$ channel yields higher branching rates than the channel with charged leptons 
in the final state. In principle, this could also be a more promising channel for probing the model in this range of masses
but requires further study.

%%-----------------------------------------------------------------
\subsection{Multi-TeV Muon Collider}
\label{sec:muon}

\begin{figure}[!t]
    \centering
    \includegraphics[width=\columnwidth]{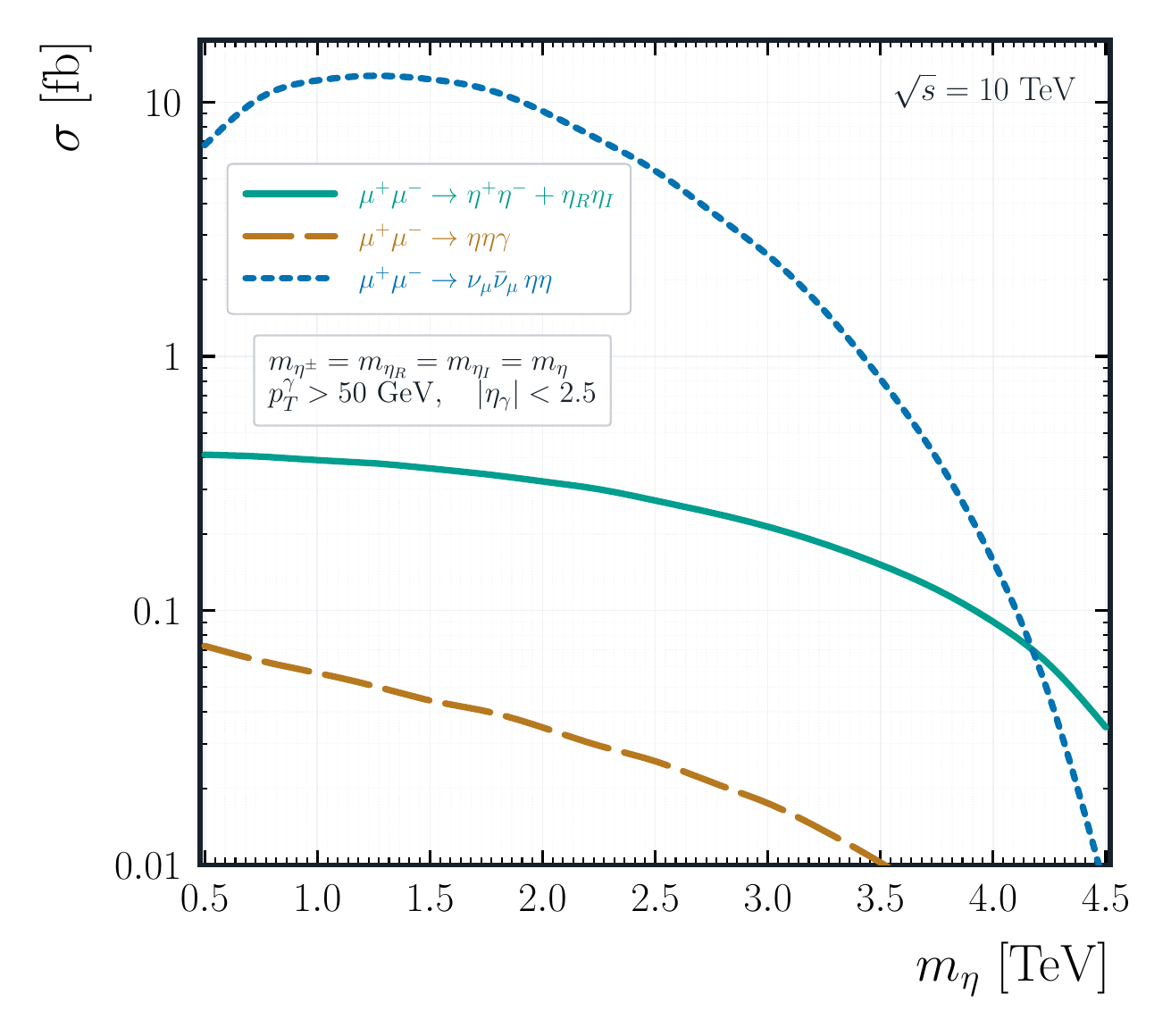}
    \caption{As a function of $m_\eta$ [TeV],
    cross sections [fb] for 
    $\eta_1\eta_2$ ({\color{teal}solid}),
    $\eta_1\eta_2\gamma$ ({\color{Goldenrod}long dash}),
    and
    $\eta_1\eta_2\nu_\mu\overline{\nu_\mu}$ ({\color{blue}short dash})
    production in $\mu^+\mu^-$ collisions at $\sqrt{s}=10\TeV$.    
    }
    \label{fig:Muon_Col}
\end{figure}

A $\mu^+\mu^-$ collider operating at a partonic center-of-mass
energy of $\sqrt{s}=10\TeV$ or higher is an exciting prospect
for the community~\cite{Black:2022cth,Accettura:2023ked,P5:2023wyd}. 
Such a facility would enable the exploration of the 
EW sector in a regime where EW bosons and top quarks
can reliably be treated as massless particles
and where weak boson content of muons can 
potentially be 
probed~\cite{Costantini:2020stv,Ruiz:2021tdt,Bigaran:2025rvb}.
Searches for new physics directly at the 
collider's center of mass energy would also be possible,
including the Scotogenic 
model~\cite{Liu:2022byu,Phan:2026yob,Mondal:2026gil}.

As a brief look towards potential opportunities 
at $\sqrt{s}=10\TeV$,
we show in Fig.~\ref{fig:Muon_Col}
the scattering rates [fb] 
as a function of scalar mass $m_\eta$ [TeV] for the 
processes
\begin{align}
    \mu^+\mu^-\ &\to\ 
    \eta_1\eta_2\ ,\ 
    \eta_1\eta_2\gamma\ ,\ \text{and}\ 
    \eta_1\eta_2\nu_\mu\bar{\nu}_\mu\ .
\end{align}
We include all interfering diagrams at LO,
except those with internal exchanges of $N_{k}$. 
This means that we consider only the impact of gauge and scalar couplings.
We otherwise keep inputs as specified in Eq.~\eqref{eq:scoto_inputs}.
For the associated-photon channel, we require
\begin{align}
    p_T^\gamma>50\GeV\ \quad\text{and}\quad \vert\eta_\gamma\vert<2.5\
\end{align}
in order to regulate infrared poles.

Over the mass ranges $m_\eta=0.5-4.5\TeV$ 
the $\eta_1\eta_2\nu_\mu\overline{\nu_\mu}$ channel ({\color{blue}short dash})
carries the largest rate, with $\sigma\sim\mathcal{O}(0.1-10)\fb$,
except when approaching half the threshold $2m_\eta\lesssim \sqrt{s}$.
Near threshold $\eta_1\eta_2$ pair production ({\color{teal}solid}) 
becomes the dominant channel with $\sigma\lesssim\mathcal{O}(0.1)\fb$.
The $\eta_1\eta_2\nu_\mu\overline{\nu_\mu}$ channel 
is driven by associated $\eta_1\eta_2 Z$ production,
including initial-state $Z$ radiation,  
and $W^+W^-$ scattering.
The dominance of these configurations highlights 
the impact of EW logarithms. 
The $\eta_1\eta_2\gamma$ channel ({\color{Goldenrod} long dash})
sits only a factor of few below the pair production channel,
similarly highlighting the impact of QED logarithms in higher leg processes.

%%-----------------------------------------------------------------
%%-----------------------------------------------------------------
\section{Conclusion}
\label{sec:conclusion}

The HL-LHC is expected to record up to
$\mathcal{L}=2-4\ab^{-1}$ of
$pp$ collisions at a
collider center-of-mass energy of
$\sqrt{s}=13.8-14\TeV$.
Such a dataset will provide an unprecedented
direct exploration of new phenomena
at the EW and TeV scales.

In this work, we have revisited the
collider phenomenology of the
Scotogenic model for neutrino masses.
The model itself is characterized by
an extended scalar sector
and singlet RH neutrinos,
both of which carry an exact $\mathbb{Z}_2$ parity.
As a result of the parity,
which is maintained
after EW symmetry breaking,
much of the model's broader phenomenology
departs from more common constructions
of the 2HDM and tree-level Seesaw models.
This is particularly true at colliders,
where the physical mass eigenstates
$\eta^\pm$, $\eta_{R/I}$, and $N_k$
must always be produced in pairs
to preserve the parity.

In Sec.~\ref{sec:theory} 
we summarized the model,
briefly listing state-of-the-art
experimental constraints.
In Sec.~\ref{sec:setup} 
we reported the development
and public release of the
{\libName} UFO libraries.
These libraries allow one
to simulate high-energy processes
in the Scotogenic model
up to NLO in QCD with parton shower matching
in software environments
commonly employed in contemporary high-energy physics.

In Sec.~\ref{sec:decay} 
we report decay rates and lifetimes
of Scotogenic particles
for various mass hierarchies.
Our ability to successfully account
for nonzero lepton masses
also provides a strong check
of the {\libName} libraries.

In Sec.~\ref{sec:production}
we present cross sections and
differential distributions
for the pair production of the scalars
$\eta^\pm$ and  $\eta_{R/I}$
through a variety of production mechanisms,
up to NLO in QCD with PS matching.
We present predictions for both
the $\sqrt{s}=14\TeV$ LHC
and a hypothetical $pp$ collider
at $\sqrt{s}=100\TeV$.
For several channels,
our numbers are the first
NLO-accurate predictions to be
reported in the literature
for the Scotogenic model.

Finally, in Sec.~\ref{sec:fccee}
we give a brief outlook for exploring
the Scotogenic model at future collider facilities.
To illustrate the possible sensitivity, 
we showed that for Scotogenic particles with
EW- and TeV-scale masses
new, rare decays of the $Z$ boson
can reach $\mathcal{O}(10^{-10})$ level,
which is well within the expected reach of the FCC-ee program.
We also showed that the production rates of TeV-scale
Scotogenic particles 
in $\sqrt{s}=10\TeV$ $\mu^+\mu^-$ collisions
surpass the fb level,
and are well within the reach 
of $\mathcal{L}\sim\mathcal{O}(1-10)\invab$ 
integrated luminosity targets.

The collider phenomenology
of the Scotogenic model
and its variants remain woefully
less studied compared
to tree-level completions
of the Weinberg operator.
We therefore hope this study
and the public release of the
{\libName} libraries
will instigate and facilitate
new theoretical studies
and experimental explorations
at the LHC.

%%-----------------------------------------------------------------
\section*{Acknowledgments}
The authors thank Marzieh Bahmani, Innes Bigaran, and Rene Poncelet 
for discussions that contributed to this work's completion.
The authors acknowledge the support of the Narodowe Centrum Nauki (NCN) under 
Grant No. 2023/49/B/ST2/04330 (SNAIL),
Grant No. 2023/49/B/ST2/03862,
and 
Grant No. 2024/55/D/ST2/00934. 
This article is based upon work from 
the COST Action COMETA (CA24146) and 
the COST Action MLQC4FC (CA24146), 
supported by COST (European Cooperation in Science \& Technology)

%%-----------------------------------------------------------------
%%-----------------------------------------------------------------

\appendix

%%-----------------------------------------------------------------
%%-----------------------------------------------------------------
\section{\texttt{SM\_Scoto} UFO Model}
\label{sec:ufo}

For the numerical calculations, we implemented the Scotogenic model 
of Ref.~\cite{Ma:2006km} into 
\texttt{FeynRules}~\cite{Christensen:2008py,Alloul:2013bka} and 
exported the libraries in the UFO format~\cite{Degrande:2011ua,Alloul:2013bka,Darme:2023jdn}.
The field conventions,
gauge quantum number and $\mathbb{Z}_2$ assignments, 
and Lagrangian are given in Sec.~\ref{sec:theory_lag}.
The UFOs are available freely from the URLs
\begin{itemize}
    \item \href{https://github.com/FeynRules/Models/tree/main/SM_Scoto}
    {https://github.com/FeynRules/Models/tree/ main/SM\_Scoto}
    \item \href{https://gitlab.cern.ch/riruiz/public-projects/-/tree/master/ScotoLHC/SM\_Scoto\_UFO}{https://gitlab.cern.ch/riruiz/public-projects/-/tree/master/ScotoLHC/SM\_Scoto\_UFO}
\end{itemize}

Several variants of the UFO versions were produced,
including those with $\mathcal{O}(\alpha_s)$ ultraviolet and $R_2$
counter terms (UFOs with \texttt{NLO} extensions) 
as well as those without (UFOs with \texttt{XLO} extensions).
The scalar potential after EW symmetry breaking is organized 
into two bases: the mass basis and the coupling basis.
In the mass basis, the
physical scalar masses are entered directly.
In the coupling basis, the
scalar-potential parameters are external inputs. 
The two versions have the same particles and vertices 
and differ only in their external inputs.
The UFOs listed in Sec.~\ref{sec:setup_sm}
and used thorughout this study 
are NLO-accurate UFOs in the mass basis.

The physical-field substitutions follow Eq.~\eqref{eq:doublets}.
The fields $\eta_R$ and $\eta_I$ and the three $N_k$ are declared
self-conjugate, while $\eta^+$ and $\eta^-$ form a particle-antiparticle
pair. The Yukawa interaction is made explicitly Hermitian before 
being combined with the SM Lagrangian:
\begin{verbatim}
	LScotYuk :=
	LYScotBase + HC[LYScotBase];
	
	LBSMScotogenic :=
	LInert + LMajorana + LScotYuk;
	
	LScotogenic :=
	LSM + LBSMScotogenic;
\end{verbatim}
Particle names and PDG codes used in the {\libName} UFO
are given in Table~\ref{tab:ufo-particles}.

\begin{table}[t]
\centering
\footnotesize
\setlength{\tabcolsep}{4pt}
\begin{ruledtabular}
    \begin{tabular}{lcc}
        State & UFO name & PDG code \\
        \hline
        $\eta_R$   & \texttt{etaR}             & $9900035$ \\
        $\eta_I$   & \texttt{etaI}             & $9900036$ \\
        $\eta^\pm$ & \texttt{eta+}, \texttt{eta-} & $\pm9900037$ \\
        $N_1$      & \texttt{n1}               & $9900012$ \\
        $N_2$      & \texttt{n2}               & $9900014$ \\
        $N_3$      & \texttt{n3}               & $9900016$
    \end{tabular}
\end{ruledtabular}
\caption{Particle names and PDG codes used in the UFO.}
	\label{tab:ufo-particles}    
\end{table}

\textbf{Mass basis:} When working in the mass basis, we use
\begin{equation}
	\left\{
	\mu_\eta^2,\lambda_2,
	m_{\eta_R},m_{\eta_I},m_{\eta^\pm}
	\right\}
	\label{eq:ufo-mass-inputs}
\end{equation}
as external inputs. The quartic couplings are then
\begin{subequations}
	\label{eq:ufo-mass-basis}
\begin{align}
	\lambda_3 &=
	\frac{2\left(m_{\eta^\pm}^2-\mu_\eta^2\right)}{v^2},
	\\
	\lambda_4 &=
	\frac{m_{\eta_R}^2+m_{\eta_I}^2
		-2m_{\eta^\pm}^2}{v^2},
	\\
	\lambda_5 &=
	\frac{m_{\eta_R}^2-m_{\eta_I}^2}{v^2}.
\end{align}
\end{subequations}
The combinations $\lambda_L$ and $\lambda_S$ then follow from their
definitions below Eq.~\eqref{eq:scalar-masses}. This basis is convenient
for collider calculations because masses are specified directly. Translating the same point between
the two bases gives the same spectrum and interaction vertices.
Once $m_{\eta^\pm}$, $m_{\eta_R}$, $m_{\eta_I}$, and
$\mu_\eta^2$ are specified, the couplings to one and two Higgs bosons
are fixed.

External parameters, the corresponding variable name,
Les Houches block assignment, and default value 
for the {\libName} in the mass basis are listed in 
Table~\ref{tab:param_values_mass_basis}.

\textbf{Coupling basis:}
When working in the coupling basis  
the external scalar inputs are
\begin{equation}
	\left\{
	\mu_\eta^2,\lambda_2,\lambda_3,\lambda_4,\lambda_5
	\right\}.
	\label{eq:ufo-coupling-inputs}
\end{equation}
The parameter $\mu_\eta^2$ is stored as \texttt{m2Eta}, while the
physical scalar masses are calculated from
Eq.~\eqref{eq:scalar-masses}. This form is useful when constraints are
applied directly to the scalar potential.

After EW symmetry breaking, the interactions of the physical
Higgs boson with $\eta^\pm$, $\eta_R$, 
and $\eta_I$ follow directly
from the scalar potential. Using
\begin{equation}
	H^0=\frac{v+h}{\sqrt{2}},
\end{equation}
and the definitions for $\lambda_L$ and $\lambda_S$ 
in Eq.~\eqref{eq:scalar-masses} 
the terms containing one or two Higgs bosons are
\begin{align}
	\mathcal{L}_{h\eta}\supset{}&
	-vh
	\left[
		\lambda_3\,\eta^+\eta^-
		+\frac{\lambda_L}{2}\eta_R^2
		+\frac{\lambda_S}{2}\eta_I^2
	\right]
	\nonumber\\
	&-\frac{h^2}{2}
	\left[
		\lambda_3\,\eta^+\eta^-
		+\frac{\lambda_L}{2}\eta_R^2
		+\frac{\lambda_S}{2}\eta_I^2
	\right].
	\label{eq:higgs-portal}
\end{align}

\begin{table}[t!]
\centering
\small
\setlength{\tabcolsep}{4pt}
\begin{tabular}{c c c c}
\hline\hline
Parameter & \texttt{FR} Name & \texttt{LH} Block & Default Value \\
\hline
$m_{\eta_R}$ 
& \texttt{MetaR} 
& \texttt{MASS} (9900035) 
& $153.9886~\mathrm{GeV}$ 
\\

$m_{\eta_I}$ 
& \texttt{MetaI} 
& \texttt{MASS} (9900036) 
& $155.9447~\mathrm{GeV}$ 
\\

$m_{\eta^\pm}$ 
& \texttt{Metap} 
& \texttt{MASS} (9900037) 
& $159.7849~\mathrm{GeV}$ 
\\

$M_{N_1}$ 
& \texttt{MN1} 
& \texttt{MASS} (9900012) 
& $300~\mathrm{GeV}$ 
\\

$M_{N_2}$ 
& \texttt{MN2} 
& \texttt{MASS} (9900014) 
& $500~\mathrm{GeV}$ 
\\

$M_{N_3}$ 
& \texttt{MN3} 
& \texttt{MASS} (9900016) 
& $1000~\mathrm{GeV}$ 
\\
\hline

$\Gamma_{\eta_R}$ 
& \texttt{WetaR} 
& \texttt{DECAY} (9900035) 
& $10.0~\mathrm{GeV}$ 
\\

$\Gamma_{\eta_I}$ 
& \texttt{WetaI} 
& \texttt{DECAY} (9900036) 
& $10.0~\mathrm{GeV}$ 
\\

$\Gamma_{\eta^\pm}$ 
& \texttt{WetaP} 
& \texttt{DECAY} (9900037) 
& $10.0~\mathrm{GeV}$ 
\\

$\Gamma_{N_1}$ 
& \texttt{WN1} 
& \texttt{DECAY} (9900012) 
& $6.3\times10^{-8}~\mathrm{GeV}$ 
\\

$\Gamma_{N_2}$ 
& \texttt{WN2} 
& \texttt{DECAY} (9900014) 
& $1.6\times10^{-7}~\mathrm{GeV}$ 
\\

$\Gamma_{N_3}$ 
& \texttt{WN3} 
& \texttt{DECAY} (9900016) 
& $3.8\times10^{-7}~\mathrm{GeV}$ 
\\
\hline

$m_\eta^2$
& \texttt{m2Eta}
& \texttt{SCOTOINERT} (1)
& $22500~\mathrm{GeV}^2$
\\

$\lambda_2$
& \texttt{lam2Eta}
& \texttt{SCOTOINERT} (2)
& $1.0$
\\
\hline

$|Y_{e1}|$
& \texttt{YeN1}
& \texttt{YSCOTO} (1)
& $1.0$
\\

$|Y_{e2}|$
& \texttt{YeN2}
& \texttt{YSCOTO} (2)
& $0.0$
\\

$|Y_{e3}|$
& \texttt{YeN3}
& \texttt{YSCOTO} (3)
& $0.0$
\\

$|Y_{\mu1}|$
& \texttt{YmuN1}
& \texttt{YSCOTO} (4)
& $0.0$
\\

$|Y_{\mu2}|$
& \texttt{YmuN2}
& \texttt{YSCOTO} (5)
& $1.0$
\\

$|Y_{\mu3}|$
& \texttt{YmuN3}
& \texttt{YSCOTO} (6)
& $0.0$
\\

$|Y_{\tau1}|$
& \texttt{YtaN1}
& \texttt{YSCOTO} (7)
& $0.0$
\\

$|Y_{\tau2}|$
& \texttt{YtaN2}
& \texttt{YSCOTO} (8)
& $0.0$
\\

$|Y_{\tau3}|$
& \texttt{YtaN3}
& \texttt{YSCOTO} (9)
& $1.0$
\\
\hline

$\delta_{e1}$
& \texttt{deN1}
& \texttt{YSCOTOPHASE} (1)
& $0.0$
\\

$\delta_{e2}$
& \texttt{deN2}
& \texttt{YSCOTOPHASE} (2)
& $0.0$
\\

$\delta_{e3}$
& \texttt{deN3}
& \texttt{YSCOTOPHASE} (3)
& $0.0$
\\

$\delta_{\mu1}$
& \texttt{dmuN1}
& \texttt{YSCOTOPHASE} (4)
& $0.0$
\\

$\delta_{\mu2}$
& \texttt{dmuN2}
& \texttt{YSCOTOPHASE} (5)
& $0.0$
\\

$\delta_{\mu3}$
& \texttt{dmuN3}
& \texttt{YSCOTOPHASE} (6)
& $0.0$
\\

$\delta_{\tau1}$
& \texttt{dtaN1}
& \texttt{YSCOTOPHASE} (7)
& $0.0$
\\

$\delta_{\tau2}$
& \texttt{dtaN2}
& \texttt{YSCOTOPHASE} (8)
& $0.0$
\\

$\delta_{\tau3}$
& \texttt{dtaN3}
& \texttt{YSCOTOPHASE} (9)
& $0.0$
\\
\hline\hline
\end{tabular}
%}
\caption{External parameters, the corresponding variable name,
the Les Houches block assignment, and their default value 
for the {\libName} UFO in the mass basis.}
\label{tab:param_values_mass_basis}
\end{table}

The first line gives the three-point couplings
$h\eta^+\eta^-$, $h\eta_R\eta_R$, and $h\eta_I\eta_I$.
The second line gives the corresponding four-point couplings
$hh\eta^+\eta^-$, $hh\eta_R\eta_R$, and $hh\eta_I\eta_I$.
For real $\lambda_5$, there are no tree-level
$h\eta_R\eta_I$ or $hh\eta_R\eta_I$ couplings.
The associated Feynman rules are
\begin{align}
	h\eta^+\eta^- &: -i\lambda_3 v,
	&
	hh\eta^+\eta^- &: -i\lambda_3,
	\nonumber\\
	h\eta_R\eta_R &: -i\lambda_L v,
	&
	hh\eta_R\eta_R &: -i\lambda_L,
	\nonumber\\
	h\eta_I\eta_I &: -i\lambda_S v,
	&
	hh\eta_I\eta_I &: -i\lambda_S.
	\label{eq:higgs-eta-vertices}
\end{align}

These couplings are fixed by the same parameters that determine the
scalar masses. In the mass basis,
\begin{subequations}
\label{eq:higgs-mass-couplings}
\begin{align}
	\lambda_3
	&=
	\frac{2\left(m_{\eta^\pm}^2-\mu_\eta^2\right)}{v^2},
\\
	\lambda_L
	&=
	\frac{2\left(m_{\eta_R}^2-\mu_\eta^2\right)}{v^2},
\\
	\lambda_S
	&=
	\frac{2\left(m_{\eta_I}^2-\mu_\eta^2\right)}{v^2}.
\end{align}
\end{subequations}

\textbf{External Parameters:}
The singlet masses $M_{N_k}$ and the nine Yukawa couplings
$Y_{\alpha k}$ are external parameters in both versions. The Yukawa
entries are stored in the \texttt{YSCOTO} block as
\texttt{YeN1-YeN3}, \texttt{YmuN1-YmuN3}, and
\texttt{YtaN1-YtaN3}. For the calculations presented here, these
parameters are real and can be varied independently. 

The particle names and model PDG codes used in the mass-basis UFO are
listed in Table~\ref{tab:ufo-particles}.

%%-----------------------------------------------------------------
%%-----------------------------------------------------------------
\section{MadGraph5\_aMC@NLO Usage}
\label{sec:mg5}

In this appendix we provide additional details 
on reproducing results reported in this work.
This appendix also provides some example usage of the {\libName}
libraries in conjunction with {\mgamc}.
Scripts used in this study are available freely from the URL
\begin{itemize}
    \item \href{https://gitlab.cern.ch/riruiz/public-projects/-/tree/master/ScotoLHC}{https://gitlab.cern.ch/riruiz/public-projects/-/tree/master/ScotoLHC}
\end{itemize}

For the $1\to2$-body decay channels 
\begin{align}
    \eta^\pm \to \ell_\alpha^\pm\ N_k\
    \quad\text{and}\quad
    \eta_{R,I} \to \nu_\alpha\ N_k\ ,
\end{align}    
we use the following {\mgamc} commands generate 
the partial decay widths given in Table~\ref{tab:scalar_decays}
\begin{verbatim}
set acknowledged_v3.1_syntax true
import model SM_Scotogenic_MassiveLeptons_4fs_NLO

generate eta+ > ta+ n1 QED=1 QCD=0
output ScotoLHC_Decay_Table_II_Hp_taN1_XLO

generate etaR > vt n1 QED=1 QCD=0
output ScotoLHC_Decay_Table_II_H0_vtN1_XLO

generate etaI > vt n1 QED=1 QCD=0
output ScotoLHC_Decay_Table_II_A0_vtN1_XLO

launch ScotoLHC_Decay_Table_II_Hp_taN1_XLO
analysis=off
set no_parton_cut
set YtaN1 1
set metap scan1:[10,100,500,1000]
set metar scan1:[10,100,500,1000]
set metai scan1:[10,100,500,1000]
set mn1   scan1:[5,5,5,5]
set nevents 40k
done
\end{verbatim}

For the NCDY channels 
in Fig.~\ref{fig:scotoLHC_xsec_vs_mass},
to generate the matrix elements at NLO in QCD
we use the syntax 
\begin{verbatim}
set acknowledged_v3.1_syntax true
import model SM_Scotogenic_NLO
define eta0 = etaI etaR

generate    p p > eta+ eta- QCD=0 QED=2 [QCD]
add process p p > eta0 eta0 QCD=0 QED=2 [QCD]
output ScotoLHC_NCDY_EtaEta_NLO_LHCX14    
\end{verbatim}
The total cross sections at this order 
are obtained using 
\begin{verbatim}
launch ScotoLHC_NCDY_EtaEta_NLO_LHCX14
order=NLO
fixed_order=ON
set metap scan1:[100, 125, 150, 175, 200, 225, 
        250, 275, 300, 325, 350, 375, 400, 
        500, 600, 700, 800, 900, 1000, 1100, 
        1200, 1300, 1400]
set metar scan1:[100, 125, 150, 175, 200, 225, 
        250, 275, 300, 325, 350, 375, 400, 
        500, 600, 700, 800, 900, 1000, 1100, 
        1200, 1300, 1400]
set metai scan1:[100, 125, 150, 175, 200, 225, 
        250, 275, 300, 325, 350, 375, 400, 
        500, 600, 700, 800, 900, 1000, 1100, 
        1200, 1300, 1400]
set req_acc_FO 0.001
set lhc 14
set pdlabel lhapdf
set lhaid 335900 # NNPDF40_nlo_as_01180_qed
set reweight_scale True
set reweight_pdf True
set dynamical_scale_choice 3
set no_parton_cut
set jetalgo -1
set jetradius 0.4
set ptj 30
set etaj 5.0
done     
\end{verbatim}
Similar commands are used to generate rates
for other processes at LO and at different energies.

For the muon collider processes in Sec.~\ref{sec:muon}
\begin{align}
    \mu^+\mu^-\ &\to\ 
    \eta_1\eta_2\ ,\ 
    \eta_1\eta_2\gamma\ ,\ \text{and}\ 
    \eta_1\eta_2\nu_\mu\bar{\nu}_\mu\ ,
\end{align}
the following generation commands were used to 
produce the matrix elements for Fig.~\ref{fig:Muon_Col}
\begin{verbatim}
generate mu+ mu- > eta+ eta- / n1 n2 n3
generate mu+ mu- > etaR etaI / n1 n2 n3

generate mu+ mu- > eta+ eta- a / n1 n2 n3
generate mu+ mu- > etaR etaI a / n1 n2 n3

generate mu+ mu- > vm vm~ eta+ eta- / n1 n2 n3
generate mu+ mu- > vm vm~ etaR etaR / n1 n2 n3
generate mu+ mu- > vm vm~ etaI etaI / n1 n2 n3
\end{verbatim}

%%-----------------------------------------------------------------
%%-----------------------------------------------------------------
%%-----------------------------------------------------------------
%%-----------------------------------------------------------------
%%-----------------------------------------------------------------

\bibliography{scotogenicLHC_refs.bib}

\end{document}